%% file: main.tex
\documentclass[preprint,12pt,authoryear]{elsarticle}

\usepackage{amssymb}
\usepackage{amsmath}
\usepackage{algorithmic}
\usepackage{textcomp}
\usepackage{multirow}
\usepackage{tcolorbox}
\usepackage{url}
\usepackage{xspace}
\usepackage{xcolor}
\usepackage{enumerate}
\usepackage{graphicx}
\usepackage{tabularx}
\usepackage{amssymb}
\usepackage{makecell}

\newcommand{\toolname}[0]{\textsc{VPFinder}\xspace}

\usepackage{ulem}

\journal{Journal of Systems and Software}

\begin{document}

\begin{frontmatter}

\title{Vulnerability Identification by Harnessing Inter-connected Multi-Source Information}

\author[Beijing]{Liyou Chen}
\ead{chenliyou@buaa.edu.cn}

\author[Beijing,Hangzhou]{Hailong Sun\corref{cor1}}
\ead{sunhl@buaa.edu.cn}

\author[Beijing,Hangzhou]{Xiang Gao\corref{cor1}}
\ead{xiang_gao@buaa.edu.cn}

\cortext[cor1]{Corresponding authors.}

\author[Beijing]{Lin Shi}
\ead{shilin@buaa.edu.cn}

\author[Beijing]{Yixin Yang}
\ead{yixinyang@buaa.edu.cn}

\author[Beijing]{Yi Xu}
\ead{yi_xu@buaa.edu.cn}

\affiliation[Beijing]{organization={State Key Laboratory of Complex \& Critical Software Environment (CCSE), Beihang University},
            city={Beijing},
            postcode={100191}, 
            country={China}}
            
\affiliation[Hangzhou]{organization={Hangzhou Innovation Institute of Beihang University},
            city={Hangzhou},
            postcode={310056},
            country={China}}

\begin{abstract}
The utilization of third-party open-source libraries is widespread in modern software development. 
Due to the dependency relationships, vulnerabilities within open-source libraries pose significant security threats to downstream software.
However, the library vulnerabilities are usually implicitly reported and patched, without explicit notification to dependent software, leaving the downstream software vulnerable to potential attacks.
Existing research efforts primarily focus on identifying vulnerability patches according to bug reports, commit messages, or code changes, overlooking the rich semantic connections among various sources of information.
In this paper, our main insight is that various sources of information, including the vulnerability descriptions (e.g., bug reports) and its fixing strategies (e.g., commit messages and code changes), are highly interconnected.
They express the high-level semantic information about the symptom, root cause and fixing strategies of the bugs.
Hence, we propose an approach that involves training an AI model to integrate multiple sources, thus enhancing the effectiveness of vulnerability identification and vulnerability type classification.
We introduce \toolname, a tool that utilizes multi-head attention mechanisms to extract high-level semantic information from diverse sources.
Evaluation results demonstrate that \toolname 
achieves remarkable 0.941 F1-score in vulnerability identification task and 0.610 F1-score in vulnerability type classification task, outperforming state-of-the-art approaches by 5.4\%.
\end{abstract}




\end{frontmatter}

\input{intro}
\input{related_work}
\input{background}
\input{approach}

\input{evaluation}

\input{discussion}
\input{conclusion}
\input{acknowledgement}

\bibliographystyle{elsarticle-harv}
\bibliography{reference}

\end{document}

%% file: intro.tex
\section{Introduction}
Open source software (OSS) plays a pivotal role in modern computing ecosystem, ranging from infrastructural software systems to application programs across various domains~\citep{blackduck2023, 11264833}.
According to Gartner~\citep{gartner}, most modern software systems are assembled rather than developed from scratch, and over 70\% of new in-house applications are expected to rely on open-source components by 2025.
Despite its benefits, OSS is frequently affected by vulnerabilities, which can lead to malicious attacks, privacy leakage, system failures, service unavailability, etc, thereby posing significant risks to downstream software systems~\citep{zhang2021investigation, shen2025understanding, 10.1145/3779222, 11185862}.

In practice, when issues are discovered in OSS projects, developers typically submit bug reports to describe, discuss, and track the corresponding problems.
These reports may involve functional bugs, feature requests, documentation issues, or security vulnerabilities.
Vulnerability-related bug reports are not always explicitly labeled, making it difficult for downstream developers to recognize potential security risks in a timely manner.
Many such reports are initially submitted to publicly accessible issue tracking systems (e.g., GitHub Issues) before being officially disclosed~\citep{li2024empirically}.
Once the repository owners are aware of a reported vulnerability, they may submit a commit to fix it or merge a pull request from the community.

Although vulnerabilities may eventually be fixed through commits or merged pull requests, such fixes are often silently patched without explicitly notifying downstream users~\citep{Wang2019Detecting,sun2023silent}.
As a a result, downstream software may remain exposed during the time window between the vulnerability fix and the official release\citep{zhang2021investigation}.
Given the vast number of issue reports and commits in modern OSS ecosystems, it is impractical for developers to manually inspect each commit to determine whether it is related to a security vulnerability, let alone assess the vulnerability's type or urgency.
With the identified vulnerability-related bug reports with patches, we can send timely security alerts based on the vulnerability's urgency for downstream software developers and encourage them to take proper actions, such as disabling the vulnerable functionality, switching to another dependency, fetching the vulnerability patch or updating dependencies.
Therefore, it is highly required to design an automated approach that can facilitate timely responses to potential security threats, thereby enhancing downstream software security and reliability. 

Existing studies have proposed automated approaches to identify vulnerability-related bug reports or patches by leveraging information from bug reports, commit messages, code changes, or combinations thereof~\citep{nguyen2022hermes, sun2023silent, 10.1145/3468854, zhou2021finding, sabetta2018practical, zhou2017automated, nguyen2022vulcurator, pan2022automated, zhang2023Howfar, Jiang2024Understanding, zhou2023colefunda, nguyen2023multi, chen2025deep, cheng2022effort}.
While these approaches have demonstrated promising results, they generally suffer from two limitations.
First, methods relying on single or limited information sources may fail to capture the rich semantic context of vulnerabilities.
Second, even approaches that utilize multiple sources often treat them independently or combine their predictions at a late stage, which limits their ability to model the intrinsic relationships between vulnerability descriptions and corresponding fixes.
Consequently, the accuracy of vulnerability identification and vulnerability type classification remains insufficient, leading to false alarms and reduced trust from downstream developers.


Similar challenges have been observed and successfully addressed in other research domains, such as multi-modal sentiment analysis~\citep{zadeh-etal-2017-tensor, tsai2019multimodal, wang2025multi}, where semantic cues are inherently distributed across heterogeneous modalities rather than being confined to a single source.
Prior studies demonstrate that explicitly modeling cross-modal semantic interactions—rather than simply concatenating features or aggregating predictions—enables models to capture fine-grained dependencies and significantly improves performance.
These findings suggest that understanding complex phenomena often requires jointly reasoning about multiple, interdependent information sources.
Inspired by this insight, our main insight is that vulnerability-related information from different sources--such as issue reports, commits messages, and code patches--is inherently interconnected, reflecting complementary perspectives of the same security problem.
Jointly modeling both the problem descriptions and their corresponding solutions can provide a more comprehensive understanding of vulnerability characteristics, root causes, and fix strategies.
However, existing multi-source approaches largely adopt implicit or shallow fusion strategies, such as feature concatenation or decision-level aggregation, which limits their ability to explicitly model fine-grained semantic interactions between vulnerability descriptions and corresponding code fixes.

In response to the above challenge, we introduce \toolname, a deep learning-based approach for identifying vulnerability patches by harnessing interconnected multi-source information.
\toolname consists of four key layers, including \textit{embedding layer}, \textit{multi-source information fusion layer} \textit{, vulnerability identification layer} and \textit{vulnerability type classification layer}.
The fusion module leverages transformer encoder~\citep{DBLP:journals/corr/abs-1810-04805, feng2020codebert} to embed both textual and code data. 
The vulnerability identification layer forecasts the relevance between the vulnerabilities and multi-source information.
If there is a correlation, \toolname further proceeds to identify its vulnerability categories; 
Otherwise, the process is terminated.
To direct the model's attention toward key vulnerability information and ignore unimportant noises, multi-source information fusion layer applies multi-head attention~\citep{10.5555/3295222.3295349} to enhance the expression of key concepts in both text and code.
In the vulnerability type classification layer, \toolname will predict the vulnerability type. 

To evaluate \toolname's effectiveness, we measure it on the dataset built on top of \textit{Pan et al.}'s dataset~\citep{pan2022automated}, PatchDB~\citep{wang2021PatchDB} and Java dataset~\citep{nguyen2023multi, zhou2021finding}.
We compare \toolname with state-of-the-art baselines: MemVul~\citep{pan2022automated}, VulFixMiner~\citep{zhou2021finding}, VulCurator~\citep{nguyen2022vulcurator}, SPI~\citep{10.1145/3468854}, \textit{Sun et al.}'s work~\citep{sun2023silent}, and TreeVul~\citep{Pan2023Fine}.
Experimental results demonstrate that the proposed approach achieves an F1-score of 0.941 in the vulnerability identification task, with a $0.061$ improvement over the best baseline MemVul~\citep{pan2022automated}.
In terms of vulnerability type classification task, \toolname achieves an F1-score of 0.610, with a $0.054$ improvement over the best baseline TreeVul~\citep{Pan2023Fine}. 
Ablation studies further validate the usefulness of each component of \toolname's architecture.
Experiments on the openEuler dataset demonstrate that certain components of \toolname also achieves promising performance in real-world scenarios.
Comparative experiments with other fusion methods (e.g., voting strategies and embedding vector concatenation) confirm the superior performance of our fusion approach.
Experimental results under varying noise ratios demonstrate that VPFinder can effectively filter out irrelevant information.
In summary, this work makes the following contributions:
\begin{itemize}
    \item We propose \toolname, a deep learning based method that identifies vulnerabilities by harnessing interconnected multi-source information. By learning and fusing the high-level semantic information from bug reports, commit messages and patches. 
    \toolname could effectively recognize vulnerabilities and their corresponding types.
    \item We evaluate \toolname and the experimental results show that
    \toolname achieves an F1 score of 0.941 in vulnerability prediction and 0.610 in vulnerability type prediction, outperforming state-of-the-art approaches.
    \item We make our dataset, tool, and model publicly accessible at: 
    \textbf{\textcolor{violet}{\url{https://github.com/KeLeXueBi/VPFinder}}}
\end{itemize}

%% file: related_work.tex
\section{RELATED WORK}
\paragraph{\textbf{Vulnerability Patch Identification.}}
To automatically identify vulnerability patches, many recent studies learn from different sources, including (1) code changes~\citep{ sabetta2018practical,Wang2019Detecting,zhou2021finding}, (2) code commit and code change~\citep{zuov2024vulnerability}, (3) bug report and commit message~\citep{zhou2017automated}, (4) bug report and code change~\citep{10.1145/3468854}.
To extract the key feature from these sources, they propose to rely on (1) different machine learning techniques~\citep{zhou2017automated, sabetta2018practical,Wang2019Detecting}, such as probability-based K-fold stacking algorithm, (2) word2vec, code2vec or transformers~\citep{zhou2021finding, sun2023silent, 10.1145/3468854} such as BERT and CodeBERT.
For instance, \textit{Wang et al}.~\citep{Wang2019Detecting} first establish a security patch database and then identify 61 code features manually, and utilize machine learning, and code clone detection to identify silent security patches.
\textit{Zhou et al.}~\citep{zhou2021finding} introduce VulFixMiner, a Transformer-based method that automatically extracts semantic meaning from commit-level code changes to identify silently patched vulnerabilities.
\textit{Nguyen et al}.~\citep{nguyen2022vulcurator, nguyen2022hermes} proposed VulCurator, leveraging deep learning on richer sources of information, including commit messages, code changes, and issue reports for vulnerability-fixing commit classification.
Although VulCurator has examined and proved the effectiveness of fusing the issue reports and patches for vulnerability identification, VulCurator only trains three classifiers for each input including the issue report, commit message, and patch, and then gets the final result with logistic regression relying on the three classifiers' outputs. While simply combining different sources is not efficient, compared to VulCurator, \toolname fuses the inputs before making a classification so \toolname is capable of uncovering the potential relationships among multi-source information.
\textit{Nguyen et al}.~\citep{nguyen2023multi} proposed MiDas. MiDas constructs different neural network models for commit-level, file-level, hunk-level, and line-level, and ultimately combines these base models to generate the final prediction.
\textit{Wang et al.}~\citep{wang2022graphspd} proposed GraphSPD, a learning-based model that represents patches as graphs with richer semantics and utilizes a patch-tailored graph model for detection. GraphSPD takes the pre-patch and post-patch source code as inputs. By merging two code property graphs for them, the model predict whether the patch is vulnerability-related or not with graph convolution and multi-layer perceptron. 
\textit{Sun et al}.~\citep{sun2023silent} propose a framework utilizing an encoder-decoder model with a binary detector to provide explainable predictions for silent dependency alerts.
The model generates key aspects, including vulnerability type, root cause, attack vector, and impact, enhancing credibility and user acceptance.
\textit{Zhou et al.}~\citep{10.1145/3468854} propose an automated security patch identification system to collect security patches in open-source software. 
The system also uses Transformer-based models and comprises two neural networks, one focusing on commit messages and the other on code changes.
Different from those works that only identify the existence of vulnerabilities, \toolname also recognizes the vulnerability type.
Moreover, \toolname fuses all the valuable information and tries to learn the latent connection among them to increase the predication accuracy.

\paragraph{\textbf{Vulnerability Type Identification.}}
To categorize vulnerability types of security patches, \textit{Pan et al}.~\citep{Pan2023Fine} firstly propose to utilize code changes to classify security patches into categories at the third level of the CWE tree.
\textit{Le et al}.~\citep{Le2021DeepCVA} introduce a deep multi-task learning model automating seven commit-level vulnerability assessment tasks based on Common Vulnerability Scoring System metrics.
\textit{Zhou et al}.~\citep{zhou2023colefunda} propose an explainable silent vulnerability fix identification framework to learn code changes and demonstrate superior performance in silent fix identification and CWE category classification.
In general, different from this kind of work, 
we not only consider the commit message and code patch but also make use of the detailed information in the bug report. This information can reflect the symptoms and root causes of the vulnerability, which can greatly improve the effect of vulnerability classification.

\paragraph{\textbf{Classifying vulnerability-related bug report.}}
Some work provides early warnings through automated identification of risky bug reports.
\textit{Peters et al}.~\citep{peters2017text} address the challenge of accurately identifying unlabelled security bug reports in bug tracking systems, and they propose a framework that filters and ranks reports to reduce the impact of security-related keywords. 
\textit{Shu et al}.~\citep{shu2021better} propose a dual optimizer to optimize models distinguishing between security bug reports and other bug reports.
\textit{Pan et al}.~\citep{pan2022automated} introduce MemVul to promptly identify bug reports that may lead to information disclosure. 
Leveraging a memory component to store vulnerability knowledge, MemVul employs an encoder, feedforward neural network, and voting mechanism to identify vulnerability-related bug reports. 
Once the vulnerability-related reports are recognized, their patches can be easily identified as vulnerability patches.
Similar to these works, \toolname can also provide early warning of vulnerabilities, although the prediction accuracy can be improved with the committed patches as inputs.  
The advantage of our work is that richer information can be used to improve the performance of vulnerability identification and vulnerability type prediction. 

\paragraph{\textbf{Positioning of \toolname}}

Existing methods such as VulCurator, MiDas, and GraphSPD leverage multiple sources of information, including commit messages, code changes, and issue reports, for vulnerability patch identification.
However, these approaches either combine sources at a late stage (e.g., VulCurator uses separate classifiers per source and merges outputs with logistic regression), or rely on specialized representations (e.g., GraphSPD constructs patch-level graphs) without explicitly modeling the semantic interactions among heterogeneous sources.
The comparison of different methods is shown in Table~\ref{tab_different_methods}.

\begin{table}[htbp]
    \centering
    \caption{Comparison of different methods.}
    \label{tab_different_methods}
    \small
    \begin{tabular}{l|cccc}
        \hline
        Method & Input Sources & Fusion Strategy & Output & Notes on Limitations \\\hline
        VulCurator & \makecell{Issue Report,\\Commit msg,\\Code Patch} & \makecell{Late fusion\\(logistic)} & \makecell{Vulnerability \\related or not} & \makecell{Ignores cross-\\modal interactions} \\\hline
        MiDas & \makecell{Commit/Line/\\File/Hunk\\levels code} & \makecell{Multi-level\\neural networks} & \makecell{Vulnerability \\related or not} & \makecell{Separate models, \\merged later} \\\hline
        GraphSPD & \makecell{Pre/post-patch\\code} & \makecell{Graph \\convolution} & \makecell{Vulnerability \\related or not} & \makecell{No textual \\info fusion} \\\hline
        \toolname & \makecell{Issue Report,\\Commit msg,\\Code Patch} & \makecell{Multi-head\\attention} & \makecell{Vulnerability\\related or not\\+\\CWE type\\(If related)} & \makecell{Early fusion, \\cross-modal \\correlations captured} \\\hline
    \end{tabular}
\end{table}

In contrast, \toolname directly fuses multiple sources through a multi-head attention mechanism, enabling the model to capture latent correlations between commit messages, bug reports, and code patches prior to classification.
This design allows \toolname to jointly consider textual descriptions and code changes, uncovering cross-modal dependencies that are ignored by prior work.
Furthermore, \toolname extends the task beyond simple vulnerability detection by also predicting the CWE type, leveraging the fused representation to improve classification accuracy.
Ablation studies confirm that both the fusion mechanism and the inclusion of all three sources contribute significantly to the overall performance, highlighting the method's novelty and practical effectiveness.

%% file: background.tex
\begin{figure}[htb]
    \centering
    \includegraphics[width=0.9\linewidth]{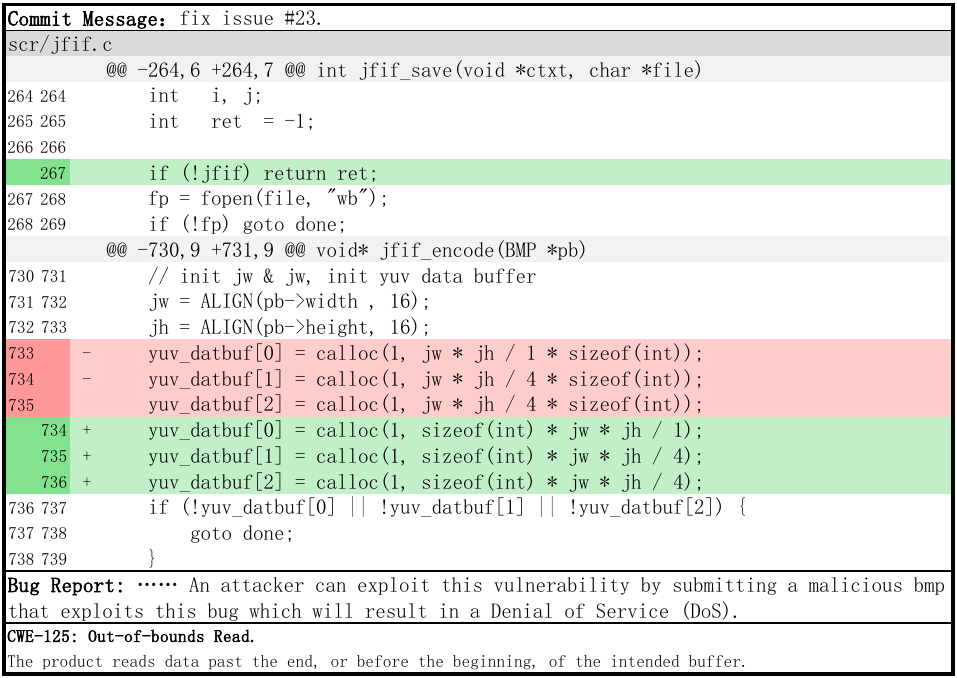}
    \caption{An example bug report and commit} 
    \label{fig:example}
\end{figure}

\begin{figure}[htb]
    \centering
    \includegraphics[width=0.9\linewidth]{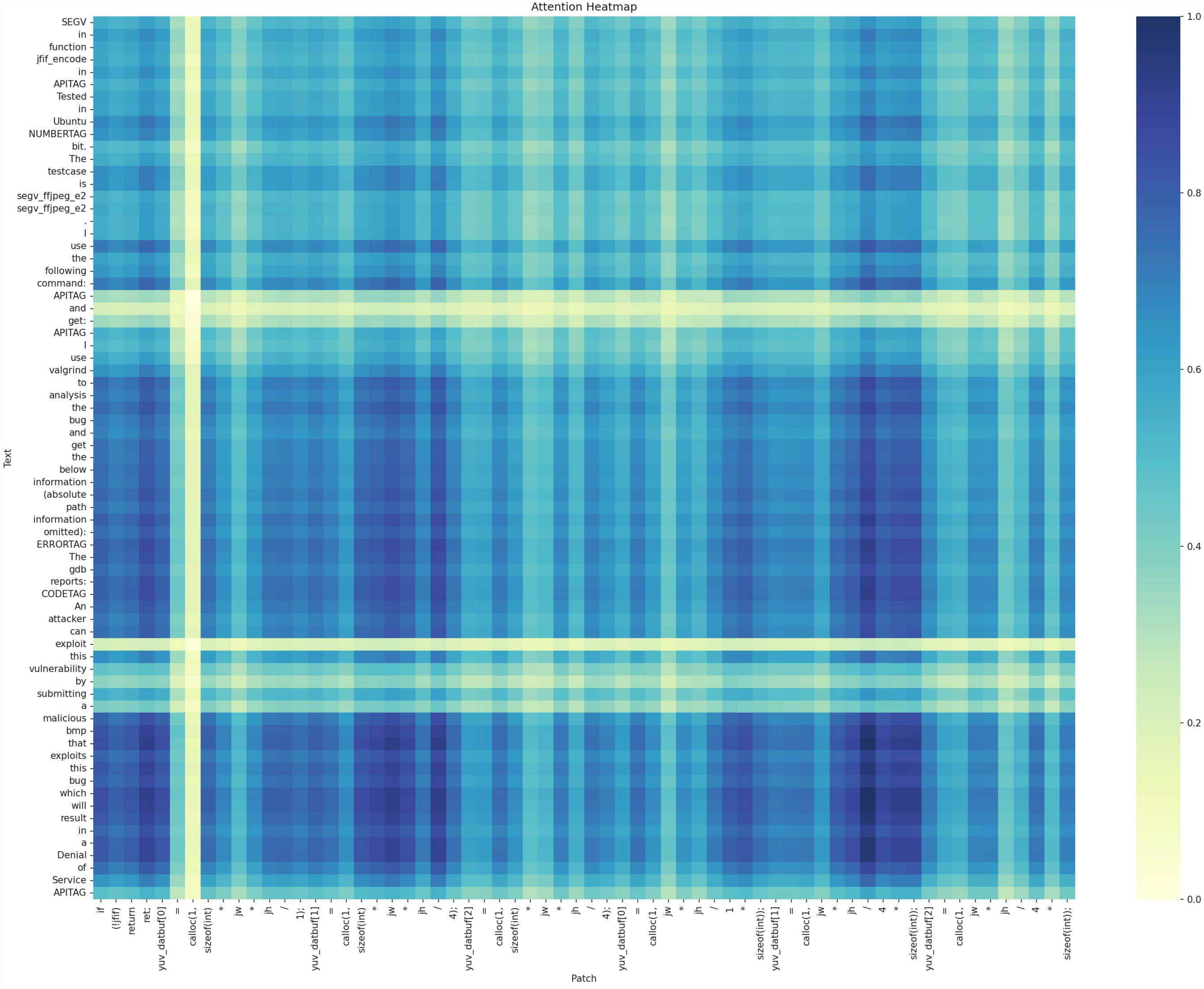} 
    \caption{Attention heatmap in \toolname.} 
    \label{fig:attention}
\end{figure}

\section{Motivating Example}
We detail our motivation and overall idea by presenting a motivating example in this section.

FFJPEG\footnote{https://github.com/rockcarry/ffjpeg} is a lightweight JPEG codec implemented purely in C, designed for embedded and resource-constrained systems. Unlike traditional libraries like libjpeg, it eliminates complex dependencies while providing core JPEG compression/decompression functionality, making it ideal for studying JPEG algorithm or optimizing compact applications. A user found a vulnerability and submitted a bug report\footnote{https://github.com/rockcarry/ffjpeg/issues/23}. The developers committed a patch to solve this vulnerability. This vulnerability corresponds to CWE-125(out-of-bound read, child of CWE-118). The key information of the report and patch is shown in Figure~\ref{fig:example}.
Specifically, the first hunk introduces a null pointer check (\textit{if (!jfif) return ret;}), which can be regarded as a robustness or defensive programming measure.
However, the bug report does not describe a null pointer dereference scenario, nor does it indicate that an attacker-controlled input could trigger such a fault.
As a result, this modification is not causally linked to the reported security impact.
In contrast, the second hunk modifies the memory allocation pattern in calloc by changing the multiplication order involving jw and jh.
This change directly mitigates an integer overflow that can lead to an undersized buffer allocation, which in turn causes out-of-bounds reads when processing a malicious BMP input.
This behavior precisely matches the exploit scenario described in the bug report (i.e., a malicious image leading to Denial of Service), and thus constitutes the actual vulnerability fix corresponding to CWE-125.
Therefore, while we acknowledge that the commit may fix multiple issues, our analysis focuses on identifying which code changes are directly related to the reported vulnerability and its type.
This distinction highlights the difficulty faced by existing tools when commits include mixed-purpose fixes, and motivates our approach to leverage bug reports and patches jointly to correctly associate vulnerability semantics with the relevant code changes.



Existing tools, that identify vulnerability patches according to the commit message and changed code, failed to identify the vulnerability in this example.
For the first patch, due to the lack of context to determine whether the ``jfif'' came from external input, there was a 60-70\% probability that they would consider it a fix for a regular bug -- a case of defensive programming~\citep{nguyen2022vulcurator, zhou2021finding}.
For the second patch, there was an 85-90\% probability they would view it as merely a bug fix~\citep{nguyen2022vulcurator, sun2023silent}, an adjustment related to code style/readability, since the total allocated memory size remained unchanged.
After referring to the bug report information, it was easy to determine that this as a vulnerability.
Besides, even if existing tools can recognize this was a vulnerability patch, it was challenging to automatically determine its vulnerability type since the bug report, commit message and patch did not mention anything about out-of-bound read.
Recognizing vulnerability type was even more important since it could give developers some hints about the severity of the vulnerability.
According to the information given in the bug report, 
we can observe this vulnerability can result in DoS. DoS attacks are typically associated with resource exhaustion (CWE-400 -- Uncontrolled Resource Consumption, belongs to the same category level of CWE-118) and insufficient input validation (CWE-20 -- Improper Input Validation, child of CWE-707).
However, the bug report included irrelevant information (Figure~\ref{fig:example} just showed partially important content), and such noise might prevent us from correctly matching the bug with the corresponding CWE. We observed existing tools, including VulCurator~\citep{nguyen2022vulcurator}, \textit{Sun et al.}'s work~\citep{sun2023silent}, and TreeVul~\citep{Pan2023Fine}, failed to recognize the correct vulnerability type for this bug.
They would consider it a null pointer dereference (according to the first patch).
However, during the execution of ``calloc'' in the second patch, the calculation method ``jw * jh / 4'' could lead to a integer overflow, resulting in a smaller buffer allocation than expected.
This could later cause out-of-bounds reads when accessing the buffer.
Modifying the ``sizeof(int)'' part might make the calculation safer by preventing the overflow or ensuring proper memory allocation.

\toolname takes various sources of information into consideration.
Specifically, \toolname relies on multi-head attention mechanisms to highlight the shared semantics between the bug report and the patch.
Figure~\ref{fig:attention} presents the normalized\footnote{During the encoding phase, we employed ``padding'' to handle insufficient data length. Since we did not prepare attention masks for the multi-head attention mechanism, the padded data inevitably consumed attention weights in the computation process. In the heatmap visualization we presented, we specifically extracted the portion of the weight matrix corresponding to the actual data length (excluding padding positions) before performing normalization.} attention heatmap from \toolname's attention mechanism, where the x-axis represents patch content (comprising both added and deleted code segments) and the y-axis corresponds to bug report content.
The visualization demonstrates that \toolname's attention focuses on two key correlations: (1) vulnerability-related descriptio\textcolor{green}{n}(``malicious bmp that exploits this bug which will result in a Denial of Service'') in the bug report, and (2) specific code modifications in the patch that address out-of-bounds read vulnerabilities, particularly the ``sizeof(int)'' and the ``jw * jh / 4'' calculation adjustment in the patch (with the vulnerability first appearing at line 734 (deleted) and line 735 (added)).
By leveraging the attention mechanism, \toolname establishes a connection between the vulnerability description in the bug report and the ``calloc'' calculation pattern in the patch.
It then predicts the vulnerability type as CWE-118: Incorrect Access of Indexable Resource (`Range Error') -- a parent category of CWE-125 (Out-of-Bounds Read).
So with the highlighted information, it can then determine whether this bug is relevant to buffer out-of-bound read, hence recognizing the correct CWE type.
In general, the bug report provide insights into the symptoms and root cause of the vulnerability, while the commit message and patch code offer details on the vulnerability's fixing strategy. 
Integrating various types of information helps \toolname effectively identify vulnerabilities and their types.


Finally, we would like to clarify that the attention weights in \toolname are not intended to indicate which specific code hunk constitutes the root-cause vulnerability fix.
Instead, they reflect the degree of semantic relevance between elements of the bug report and different parts of the patch.
In this example, although the first hunk does not address the root cause of the out-of-bounds read, it introduces a defensive null check that is semantically related to the vulnerability symptoms described in the bug report, such as program crashes and denial-of-service behavior.
This explains why the attention mechanism assigns non-negligible weights to the first hunk.
Importantly, the actual root cause of the vulnerability—an integer overflow in the buffer size computation leading to out-of-bounds reads—is only addressed in the second hunk.
While both hunks are semantically relevant at different levels, the second hunk provides the critical information required for inferring the correct vulnerability type.
\toolname aggregates information across all hunks and modalities, and the final prediction is not determined by attention to a single patch fragment.
Consequently, the presence of attention on the first hunk does not indicate misclassification, but rather reflects the model’s ability to capture contextual and symptom-level relevance in addition to root-cause signals.
This behavior is consistent with real-world development practices, where vulnerability-fixing commits often include auxiliary defensive changes alongside the actual root-cause fix, and we believe it demonstrates the advantage of modeling cross-modal semantic relationships rather than relying on isolated code patterns.

%% file: approach.tex
\section{APPROACH}\label{sec:approach}

To identify vulnerabilities and their type, we propose a deep learning based approach \toolname.
The inputs of \toolname include two main parts: (1) \textbf{bug description}: the problem description from the bug report; (2) \textbf{bug fix}: the commit message and the patch.
The output is whether the report or patch is related to vulnerability, and vulnerability type (if applicable).
We first present the overall architecture of \toolname, and then detail each step.

\subsection{Overall Architecture}

\begin{figure}[htb]
  \centering
   \includegraphics[scale=0.4]{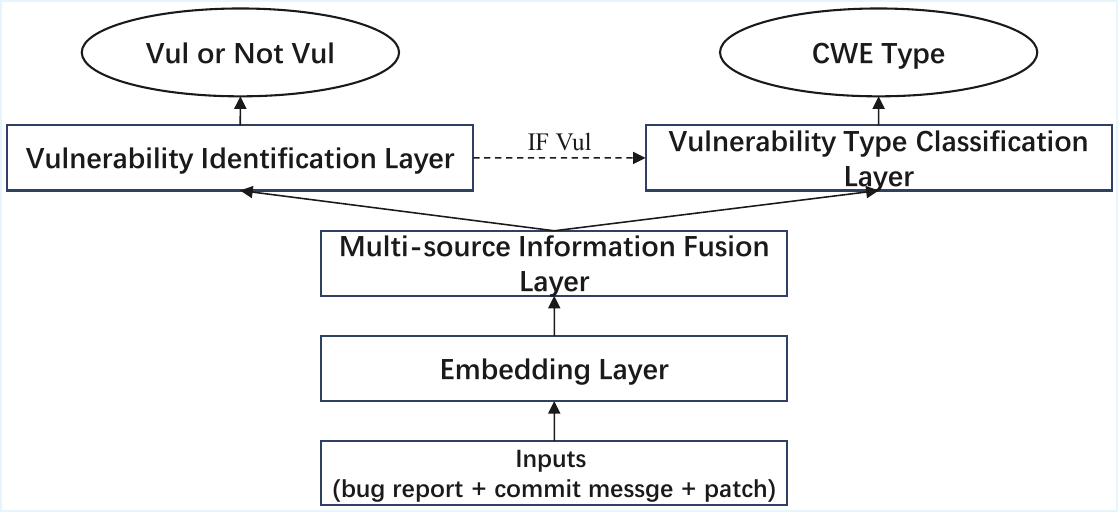}
  \caption{The overall architecture of \toolname.}
  \label{fig:arch}
\end{figure}

The overall structure of \toolname is illustrated in Figure~\ref{fig:arch}, including four main layers: (1) embedding layer, (2) multi-source information fusion layer, (3) vulnerability identification layer, and (4)vulnerability type classification layer. 
First, \toolname takes bug description, bug fix and patch as inputs, which are then fused together to better represent the features of the bug in the embedding layer.
Specifically, \toolname utilizes two encoders to learn the vector representations of the text and code respectively.
Second, in the multi-source information fusion layer, \toolname employs multi-head attention and residual blocks~\citep{He2015} to capture high-level semantic relationships among multi-source information. In the vulnerability identification layer, \toolname integrates the fusion results with preliminary classification outputs from the embedding layer to predict the relevance between multi-source information and vulnerabilities. If the bug is identified as a vulnerability in vulnerability identification layer, the vulnerability type classification layer will continue classifying the associated CWE types; otherwise, the process concludes.



\begin{figure}[htb]
  \centering
  \includegraphics[width=\linewidth]{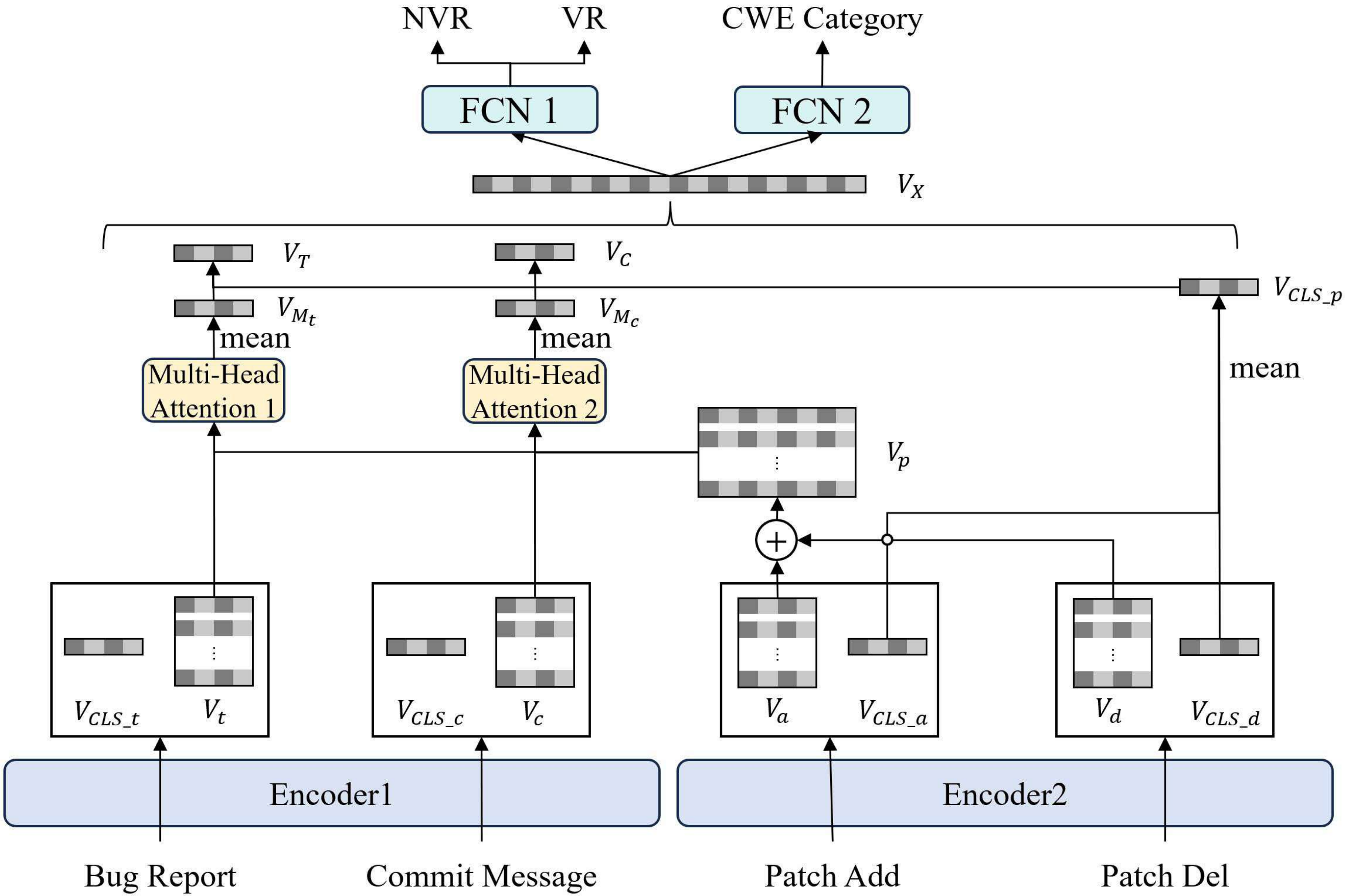}
  \caption{The overall framework of \toolname.}
  \label{fig:model}
\end{figure}

\subsection{Embedding Layer}

The fusion of multi-source information dedicates to fusing the high-level semantic bug information from multiple sources.
Given $N$ samples in the dataset, as shown in Equations (\ref{Eq 1})-(\ref{Eq 4}),
this module takes the problem description \(X_{description_{i}}\), commit message \(X_{message_{i}}\) and patch \(X_{patch_{i}}\) as inputs, 
for \(0 < i <= N\).
Patch vector \(X_{patch_{i}}\) consists of two parts: the added code \(X_{patch\ add_{i}}\) and the deleted code \(X_{patch\ del_{i}}\).
We encode the text-based inputs such as problem descriptions and commit messages with \texttt{Encoder1}, and encode code-based inputs such as  patch with \texttt{Encoder2}.
\texttt{Encoder1} can be any tool like BERT that can process text information and \texttt{Encoder2} can be any tool like CodeBert that can process code and annotation information.
\begin{equation}
    (V_{t_{i}}, V_{CLS\_t_{i}}) = f_{Encoder1}\left ( X_{description_{i}} \right )
    \label{Eq 1}
\end{equation}
\begin{equation}
    (V_{c_{i}}, V_{CLS\_c_{i}}) = f_{Encoder1}\left ( X_{commit\ msg_{i}} \right )
    \label{Eq 2}
\end{equation}
\begin{equation}
    (V_{a_{i}}, V_{CLS\_a_{i}}) = f_{Encoder2}\left ( X_{patch\ add_{i}} \right )
    \label{Eq 3}
\end{equation}
\begin{equation}
    (V_{d_{i}}, V_{CLS\_d_{i}}) = f_{Encoder2}\left ( X_{patch\ del_{i}} \right )
    \label{Eq 4}
\end{equation}
The problem descriptions are embedded to yield matrix: \((V_{t_{i}}, V_{CLS\_t_{i}})\), where \(V_{t_{i}}\) denotes the hidden states of \texttt{Encoder1} at each position, and \(V_{CLS\_t_{i}}\) is a single vector representing the entire text.
Similarly, \texttt{Encoder1} is also applied to commit messages. 
Moreover, 
\texttt{Encoder2} is applied to yield patch representation. 
The [CLS] token of each encoder output has rich semantic information so it is capable of vulnerability identification by simply fusing.
However, for vulnerability type classification, just using tokens is not enough.
The rich semantic information needs further fusion to establish potential links between these types of information.
We'll discuss how to fuse multi-source information using multi-head attention and residual blocks.

\subsection{Multi-Source Information Fusion Layer}

As shown in Figure~\ref{fig:model}, multi-source information plays a crucial role in \toolname, as it helps to better understand and address the vulnerabilities in each component.
\toolname utilizes multi-source information layer to establish potential connections between the multi-source information.
Because bug report and commit message have a certain connection with patches, hidden states will be further used to explore the potential connection between them.

\subsubsection{Processing of patches}

\toolname first concatenates the hidden states of the added code and the deleted code, as shown in Equation (\ref{Eq 10}).
The hidden states \(V_{p_{i}}\) has semantic information of the patch.

\begin{equation}
    V_{p_{i}} = concat(V_{a_{i}}, V_{d_{i}})
    \label{Eq 10}
\end{equation}

Then \toolname takes the arithmetic mean of the added code and deleted code to obtain the semantic information of the patch, as shown in Equation (\ref{Eq 11}).
The obtained average patch semantic representation will be used in the residual blocks~\citep{He2015} to prevent gradient vanishing and gradient explosion.

\begin{equation}
    V_{CLS\_p_{i}} = \frac{1}{2}(V_{CLS\_a_{i}} + V_{CLS\_d_{i}})
    \label{Eq 11}
\end{equation}

\subsubsection{Multi-Head Attention Layer}

The bug report and commit message may include a large part of the irrelevant content to the patch, which may significantly affect the vulnerability identification and classification accuracy.
To solve this problem, \toolname applies attention mechanism~\citep{10.5555/3295222.3295349} to highlight the patch content by measuring the shared semantic similarities between them.
The multi-head attention mechanism is built on the basis of self-attention~\citep{10.5555/3295222.3295349}, and captures different aspects of information in the input sequence by computing multiple attention heads in parallel.
Each head focuses on a different subspace of the input data, which helps the model understand the input information from multiple perspectives and improves the model's ability to express and generalize.
The results of all heads are then spliced and then subjected to linear transformation to obtain the final output.
Specifically, relying on an attention mechanism, we propose to treat the hidden state of bug report \(V_{t_{i}}\) as \(\mathit{Query}\), and the information at each position of the patch sequence as \(\mathit{Key}\) and \(\mathit{Value}\).
As shown in Equation (\ref{Eq 12}), \(\textup{Q}\), \(\textup{K}\), and \(\textup{V}\) represent \(\mathit{Query}\), \(\mathit{Key}\), and \(\mathit{Value}\), respectively.

\begin{equation}
    \mathit{Attention}\left ( \textup{Q} {=} V_{t_{i}}, \textup{K} {=} V_{p_{i}}, \textup{V} {=} V_{p_{i}} \right ) {=} \mathit{softmax}\left ( \frac{\textup{Q} 
\cdot \textup{K}^{T}} {\sqrt{d_{k}}} \right ) \textup{V}
    \label{Eq 12}
\end{equation}

\(\mathit{Query}\) and \(\mathit{Key}\) are dot multiplied to obtain the similarity between them. 
The similarity is then scaled using $\sqrt{d_k}$, where $d_k$ is the dimensions of the \(\mathit{Key}\) vector. 
Then, the \(softmax\) operation is performed to convert the similarity into the weight of \([0, 1]\) interval. 
A higher similarity between the \(\mathit{Query}\) and the \(\mathit{Key}\) results in a higher attention weight.
Finally, the weight is multiplied by the \(\mathit{Value}\) to obtain the weighted results.
In this process, the parts with similar semantics between buggy code and problem description will receive more ``attention''.

\toolname then trains a multi-head attention~\citep{10.5555/3295222.3295349}, as shown in Equation (\ref{Eq 13}), where \( W_{j}^{\textup{Q}}\), \( W_{j}^{\textup{K}}\), \( W_{j}^{\textup{V}}\) and \( W^{\textup{O}}\) are trainable parameters. \(j \in (0, J)\) represents the \(j\)-th head and \(J\) represents the number of heads.
\begin{equation}
\begin{gathered}
MultiHead\left ( \textup{Q}, \textup{K}, \textup{V} \right ) = concat\left ( head_{1}, ... , head_{J} \right ) W^{\textup{O}}\\
\texttt{\textbf{where}}\ head_{j} {=} Attention\left ( \textup{Q} \cdot W_{j}^{\textup{Q}}, \textup{K} \cdot W_{j}^{\textup{K}}, \textup{V} \cdot W_{j}^{\textup{V}} \right )
\end{gathered}
    \label{Eq 13}
\end{equation}
    


Specifically, multi-head attention divides the input into multiple heads, with each head learning different weights.
As shown in Equation (\ref{Eq 13}), in multi-head attention, each attention head has its own set of weight matrices $W_{j}^{\textup{Q}}$, $W_{j}^{\textup{K}}$ and $W_{j}^{\textup{V}}$ designed to associate different parts of the input sequence. 
Each head attention score is then calculated using Equation (\ref{Eq 12}) for training stability. 
Finally, all head outputs are concatenated, forming the output of the multi-head attention. Further matrix multiplication is applied through a linear projection matrix $W^{\textup{O}}$ to obtain the final output of the multi-head attention. 
Linear projection allows the model to better capture information from different positions when dealing with sequential data, fusing both performance and generalization capabilities.
To fuse bug report and patch, we calculate multi-head attention $M_{t_{i}} = MultiHead \left ( V_{t_{i}}, V_{p_{i}}, V_{p_{i}} \right )$, which corresponds to \texttt{Multi-Head Attention 1} in Figure~\ref{fig:model}.
By calculating attention scores for $V_{p_{i}}$ based on the $V_{t_{i}}$, and then performing a weighted sum on $V_{p_{i}}$, the semantics related to vulnerabilities are enhanced. 
The output is matrix $M_t$, where each row represents the semantic information of the patch sequence at each position.
Average pooling helps in extracting global information from the entire sequence and reducing dimensionality, aiding the model in learning the overall features of the sequence.
So an average pooling is performed on the output matrix to obtain a smooth semantic representation of the patch, as shown in Equation (\ref{Eq 14})-(\ref{Eq 15}), where $M_{t_{i_{jk}}}$ represents the element in the $k$-th row and $j$-th column of matrix $M_{t_{i}}$, and m represents the number of rows in matrix $M_{t_{i}}$.

\begin{equation}
    \bar{V}_{M_{t_{i}}} = \frac{1}{m} \sum_{j=1}^{m} M_{t_{i_{jk}}}
    \label{Eq 14}
\end{equation}
\begin{equation}
    V_{T_{i}} = concat( \bar{V}_{M_{t_{i_{1}}}},...,\bar{V}_{M_{t_{i_{m}}}} )
    \label{Eq 15}
\end{equation}

Moreover, \toolname also applies the same attention approach to fuse the commit message information with the similarity between commit message and patch. 
Specifically, we calculate the multi-head attention $M_{c_{i}} = MultiHead \left ( V_{c_{i}}, V_{p_{i}}, V_{p_{i}} \right )$, corresponding to \texttt{Multi-Head Attention 2} in Figure~\ref{fig:model}.
In the same way, \toolname also utilizes average pooling to obtain the smooth semantic representation of the patch, as shown in Equation (\ref{Eq 16})-(\ref{Eq 17}).

\begin{equation}
    \bar{V}_{M_{c_{i}}} = \frac{1}{m} \sum_{j=1}^{m} M_{c_{i_{jk}}}
    \label{Eq 16}
\end{equation}
\begin{equation}
    V_{C_{i}} = concat( \bar{V}_{M_{c_{i_{1}}}},...,\bar{V}_{M_{t_{i_{m}}}} )
    \label{Eq 17}
\end{equation}

\subsubsection{Residual Block}

Residual blocks~\citep{He2015}, a cornerstone of modern convolutional neural network (CNN) architectures, have emerged as a pivotal solution to the long-standing issue of training very deep networks.
These blocks incorporate shortcut connections that enable the input of a block to be directly added to its output, bypassing the series of convolutions and non-linear transformations within the block.
This design principle, known as residual learning, addresses the degradation problem encountered in traditional deep networks, where the addition of layers often leads to higher training error.
By facilitating the flow of gradients during backpropagation and enabling features from earlier layers to propagate deeper into the network, residual blocks enable the construction of networks with unprecedented depths, thereby pushing the boundaries of what is achievable in computer vision tasks such as image classification, object detection, and semantic segmentation.
The effectiveness of residual blocks has been amply demonstrated through their widespread adoption in state-of-the-art CNN architectures~\citep{li2024lors}, including the seminal ResNet family, which has inspired numerous follow-up works~\citep{he2016deep, wang2017residual} that further explore and refine the concept of residual learning.

Through the skip connection in residual block, the features of the previous layer can be directly used by the subsequent layers, which helps the network to share and reuse features between different layers and improves the efficiency of the network.
This idea can also be used in \toolname, as shown in Equation (\ref{Eq 18})-(\ref{Eq 19}).
Residual blocks helps reuse features and increase depth and complexity of the network.

\begin{equation}
    V_{T_{i}}^{'} = V_{T_{i}} + V_{CLS\_p_{i}}
    \label{Eq 18}
\end{equation}
\begin{equation}
    V_{C_{i}}^{'} = V_{C_{i}} + V_{CLS\_p_{i}}
    \label{Eq 19}
\end{equation}

Finally, \toolname concatenates the residual blocks' outputs and all [CLS] token vectors, as shown in Equation (\ref{Eq 20}).

\begin{equation}
    V_{X_{i}} = cancat(V_{T_{i}}^{'}, V_{C_{i}}^{'}, V_{CLS\_t_{i}}, V_{CLS\_c_{i}}, V_{CLS\_p_{i}})
    \label{Eq 20}
\end{equation}

Up to this point, we have completed the fusion of multi-source information, obtaining the final vector representation $V_{X_{i}}$.

\subsection{Vulnerability Identification Layer}

As shown in Fig~\ref{fig:model}, to identify whether the bug report with relevant commits are related to vulnerability, \toolname concatenates each [CLS] token output by BERT or CodeBERT and utilizes a fully-connected neural network \(FCN_{1}\).
\(FCN_{1}\)'s output is the probability (\(P_{vul_{i}}\)) representing whether this bug is associated with vulnerability.
The above process is shown in Equation (\ref{Eq 24}).

\begin{equation}
    P_{vul_{i}} = FCN_{1}(V_{X_{i}})
    \label{Eq 24}
\end{equation}

Then \toolname generates the prediction result via Equation (\ref{Eq 25}), where \(\tilde{z}_{i}\) denotes the label of the \(i\)-th sample predicted by the model and \(\tilde{z}_{i} = 1\) indicates that this sample is predicted to be related to the vulnerability.

\begin{equation}
    \tilde{z}_{i} = \mathop{\arg \max} P_{vul_{i}}
    \label{Eq 25}
\end{equation}

To learn the involved parameters, the cross-entropy loss function is applied, as demonstrated in Equation (\ref{Eq 26}), where \(z_{i}\) denotes the ground truth of \(i\)-th sample and \(z_{i} = 1\) indicates that this bug is related to vulnerability.
\begin{equation}
    \mathfrak{L}_{1} = -\frac{1}{N} \sum_{i=1}^{N} \left ( z_{i} \log \tilde{z}_{i} + \left ( 1 - z_{i} \right ) \log \left ( 1 - \tilde{z}_{i} \right ) \right )
    \label{Eq 26}
\end{equation}

To learn the involved parameters, the cross-entropy loss function is applied, as demonstrated in Equation (\ref{Eq 27}), where \(z_{i}\) denotes the ground truth of \(i\)-th sample and \(z_{i} = 1\) indicates that this bug is related to vulnerability.
\begin{equation}
    \mathfrak{L}_{1} = -\frac{1}{N} \sum_{i=1}^{N} \left ( z_{i} \log \tilde{z}_{i} + \left ( 1 - z_{i} \right ) \log \left ( 1 - \tilde{z}_{i} \right ) \right )
    \label{Eq 27}
\end{equation}

If a bug report with relevant commits are predicted as vulnerability-related, \toolname will further predict its vulnerability type.

\subsection{Vulnerability Type Classification Layer}
\toolname utilizes another fully connected neural network $FCN_{2}$ to extract features and outputs the probability that the vulnerability corresponds to each categories, as shown in Equation (\ref{Eq 21}).

\begin{equation}
    P_{category_{i}} = FCN_{2}(V_{X_{i}})
    \label{Eq 21}
\end{equation}

Then, \toolname generates the classification results by Equation (\ref{Eq 22}), where \(\tilde{r}_{i}\) denotes the label of the \(i\)-th sample predicted by the model and \(\tilde{r}_{i} = k\) indicates that this bug is predicted to be the \(k\)-th category of CWE. 
\begin{equation}
    \tilde{r}_{i} = \mathop{\arg \max} P_{class_{i}}
    \label{Eq 22}
\end{equation}

Similar to vulnerability identification task, to learn the involved parameters, we utilize the cross-entropy loss function, as depicted in Equation (\ref{Eq 23}), where \(r_{i}\) denotes the ground truth of \(i\)-th sample and \(r_{i} = k\) indicates that this bug belongs to \(k\)-th category of CWE.
\begin{equation}
    \mathfrak{L}_{2} = -\frac{1}{N} \sum_{i=1}^{N} \left ( r_{i} \log \tilde{r}_{i} + \left ( 1 - r_{i} \right ) \log \left ( 1 - \tilde{r}_{i} \right ) \right )
    \label{Eq 23}
\end{equation}

%% file: evaluation.tex
\section{EXPERIMENT}

In this section, we describe our research questions, dataset, experiment setting and results in detail. 
Specifically, we evaluate \toolname to answer the following research questions:

\begin{itemize}
    \item \textbf{RQ1:} How effective is \toolname in identifying vulnerabilities based on multi-source information?
    \item \textbf{RQ2:} How effective is \toolname in classifying vulnerability types based on multi-source information?
    \item \textbf{RQ3:} How do architectural variations influence the performance of \toolname?
    \item \textbf{RQ4:} How effective is \toolname in real application scenarios on key software package repositories of openEuler~\citep{openEuler} dependencies?
    \item \textbf{RQ5:} Does \toolname's ability to filter out irrelevant information outperform other methods?
\end{itemize}

\subsection{Dataset}\label{sec: data}

To evaluate \toolname, we directly reuse the MemVul dataset provided by \textit{Pan et al. }~\citep{pan2022automated}.
MemVul dataset includes 3884 CVE-referred issue reports and nearly 1191\textbf{$k$} not CVE-referred issue reports.
For each sample in this dataset, we extract its URL, problem description, and other vulnerability information. 
Specifically, we utilize GitHub's REST API~\citep{github_api} to access the issue reports and relevant commits in GitHub repositories.
We make an effort to find the commit associated with each issue and record the SHA value of the relevant commit.
If we cannot find related commit information in the bug report, we proceed to check the discussion section of the bug report.
If the commits are not available, the bug report will not be included in our dataset.
Once we identify the relevant commits, we use the REST API to retrieve the patch information for the commit.
Further, we extract the modified code (including added and deleted code) based on the commit.
If more than one patch file exists, we merge them together.
We just retain C, Java, and Python projects, and filter out samples whose fixes do not involve modifying code.
After filtering out patches and limiting the dataset to maintain a roughly 1:2 ratio of vulnerability-fixing samples to non-vulnerability-fixing samples, the final dataset contains 1,090 vulnerability-fixing and 3,385 non-vulnerability-fixing samples.
The enhanced MemVul dataset that we used contains all the bug information, including flag (i.e., vulnerability identification label), text (i.e., bug description), message (i.e., commit message), patch, CVE\_ID, and CWE\_ID.
Based on the year disclosed in the CVE ID (vulnerability-related samples) or the year that bug report created at (non-vulnerability-related samples), we categorize all samples into training set (prior to 2020) and testing set (2020).

To observe the performance of \toolname in more real-world datasets, we also compare \toolname with existing tools on PatchDB~\citep{wang2021PatchDB}, and Java dataset~\citep{nguyen2023multi, zhou2021finding}.
PatchDB is a large-scale security patch dataset that contains around 12K security patches and 24K non-security patches from the real world.
We collected the associated issues if available.
Java dataset contains vulnerability fixing and non-vulnerability fixing commits collected from the real world.
Table \ref{dataset} presents the statistics of the dataset.
The experiments conducted on the PatchDB and Java datasets are performed using the complete datasets, and we adhere to the datasets' original training and testing splits.

\begin{table}[htb]
    \centering
    \caption{The number of samples in the dataset.}
    \begin{tabular}{l|l r r r}
        \hline
        Dataset & Type & Vul & NVul & Total \\\hline
        \multirow{2}{*}{MemVul} & Training set & 961 & 2041 & 3002 \\
        & Testing set & 129 & 251 & 380 \\
        \hline

        \multirow{2}{*}{PatchDB} & Training set & 9658 & 18993 & 28651 \\
        & Testing set & 2415 & 4749 & 7164 \\
        \hline

        \multirow{2}{*}{Java} & Training set & 983 & 31323 & 32306 \\
        & Testing set & 300 & 87856 & 88156 \\
        \hline
    \end{tabular}
    \label{dataset}
\end{table}

For vulnerability type classification, we observe that the number of samples belonging to some CWE categories is too low.
Hence, we merge the less frequent CWE categories in the dataset into composite categories.
When merging minority classes, we follow the tree structure~\citep{mitre_cwe}, combining minority classes belonging to the same parent node into the parent class.
For example, CWE-693, CWE-435, CWE-697, and CWE-703 belong to the first layer in the CWE tree, so we combine these CWE samples as `CWE-1'. In addition, CWE-417 is a CWE category (communication channel errors) although it has been discouraged since 2019, so we combine it into CWE-1, too. CWE-913, CWE-706, CWE-704, CWE-669, CWE-666, and CWE-665 belong to the second layer and they are all child nodes of CWE-664, so we combine them as `CWE-2'.
However, there are still some classes with too few samples, and we merge them into new classes.
Additionally, for individual samples with undetermined CWE, we categorize these samples as `CWE-1000'.
Due to limitations in sample quantity, we chose to provide the second-layer CWE category only for `CWE-664'.
This results in eight labels at the first layer and four labels at the second layer.
The CWE statistics are detailed in Table \ref{CWE statistics}.
The labels are exclusive, indicating that if a sample is predicted as its parent node's type, it is considered as a misprediction.

For we sacrifice some CWE granularity to gain statistical reliability, and the resulting coarse-grained classification remains meaningful and practically relevant.
Merging CWE categories enables statistically reliable training and evaluation.
Without this step, the performance on rare CWE classes would be dominated by noise, making it difficult to draw meaningful conclusions about model effectiveness.
Importantly, while the merging operation results in a coarser-grained classification setting, the task remains meaningful for our study objectives.
The merged CWE categories correspond to semantically related vulnerability families, allowing the model to capture higher-level vulnerability patterns rather than overly specific instances.
As a result, the obtained results still provide valid insights into the model’s ability to distinguish different classes of vulnerabilities at the family level.

\begin{table}[htb]
    \centering
    \caption{Dataset CWE statistics.}
    \begin{tabular}{l|l|c c c}
        \hline
         & CWE category & Number & Label \\\hline
        \multirow{8}*{First layer} & CWE-664 & 156 & 1 \\
         & CWE-707 & 90 & 2 \\
         & CWE-710 & 83 & 3 \\
         & CWE-682 & 57 & 4 \\
         & CWE-691 & 58 & 5 \\
         & CWE-1(merged category) & 27 & 6 \\
         & CWE-284 & 33 & 7 \\
         & CWE-1000 & 26 & 8 \\\hline
        \multirow{4}*{Second layer} & CWE-118 & 398 & 9 \\
         & CWE-404 & 75 & 10 \\
         & CWE-668 & 47 & 11 \\
         & CWE-2(merged category) & 40 & 12 \\\hline 
    \end{tabular}
    \label{CWE statistics}
\end{table}

\subsection{Experiment Setting}

\subsubsection{Baselines}
We compare \toolname with state-of-the-art work, including MemVul~\citep{pan2022automated}, VulFixMiner~\citep{zhou2021finding}, GraphSPD~\citep{wang2022graphspd}, MiDas~\citep{nguyen2023multi}, VulCurator~\citep{nguyen2022vulcurator}, SPI~\citep{10.1145/3468854}, \textit{Sun et al.}'s work~\citep{sun2023silent}, and Treevul~\citep{Pan2023Fine}.
Specifically,
\begin{itemize}
    \item \textbf{MemVul:} MemVul utilizes language model and memory component to predict whether a bug report is related to a vulnerability.
    \item \textbf{VulFixMiner:} VulFixMiner is a Transformer-based model for identifying cross-project and cross-language vulnerability fixes according to code changes. 
    \item \textbf{GraphSPD:} GraphSPD represents patches as graphs with richer semantics and utilizes a patch-tailored graph model for identifying vulnerability patches.

    \item \textbf{VulCurator:} VulCurator trains classifiers for bug reports, commit messages, and patches, and combines the outputs of the three classifiers using logistic regression to identify vulnerabilities. Compared to \toolname fusing the information at the embedding-level, VulCurator fuses the information at the decision-level.

    \item \textbf{SPI:} SPI learns features from commit messages and patch, and selects high-level features for an information-rich, latent semantic representation. Then SPI effectively integrates the unified representation to build a classifier to determine whether a commit is a security patch or not.
    
    \item \textbf{\textit{Sun et al.}'s work:} This model utilizes an encoder-decoder framework to identify vulnerability patches according to commit message and code changes.

    \item \textbf{MiDas:} MiDas constructs different base models for each level of code change granularity, corresponding to commit-level, file-level, hunk-level, and line-level, and combines all base models to output the final prediction. As same as VulCurator, MiDas fuses the information at the decision-level. MiDas also incorporates feature fusion, while its fusion focuses on how to represent contextual information. For instance, it considers whether to treat the contextual information as a whole for internal fusion within the context or to concatenate the added and removed code into a vector for internal fusion. Both of these fusion approaches do not involve the integration of multi-source information.
    
    \item \textbf{TreeVul:} TreeVul takes as inputs commit message and changed code, and classifies vulnerability patches via a hierarchical and chained architecture.
\end{itemize}
Among them, MemVul, VulFixMiner, SPI, \textit{Sun et al.}'s work, VulCurator, GraphSPD, and MiDas are used for comparison on the vulnerability identification task.
\textit{Sun et al.}'s work and TreeVul are used for comparison on the vulnerability type classification task.
In addition, we also extended VulCurator to support vulnerability type classification task to evaluate the performance of decision-level fusion.
Table~\ref{tab:model_types} shows the input types used by each model and the types of tasks they can perform.

\begin{table}[htbp]
    \centering
    \small
    \caption{Comparison of models.}
    \label{tab:model_types}
    \begin{tabularx}{\textwidth}{|p{2cm}|p{1.3cm}|p{1.4cm}|p{1cm}|p{2.2cm}|X|}
    \hline
        Model & \multicolumn{3}{c|}{Inputs} & \multicolumn{2}{c|}{Tasks} \\ \hline
        ~ & Issue Report & Commit Message & Code Patch & vulnerability identification & vulnerability type classification \\ \hline
        MemVul & \checkmark & ~ & ~ & \checkmark & ~ \\ \hline
        VulFixMiner & ~ & ~ & \checkmark & \checkmark & ~ \\ \hline
        VulCurator & \checkmark & \checkmark & \checkmark & \checkmark & \checkmark \\ \hline
        SPI & ~ & ~ & \checkmark & \checkmark & ~ \\ \hline
        \textit{Sun et al}.'s work & ~ & \checkmark & \checkmark & \checkmark & \checkmark \\ \hline
        GraphSPD & ~ & ~ & \checkmark & \checkmark & ~ \\ \hline
        MiDas & ~ & ~ & \checkmark & \checkmark & ~ \\ \hline
        TreeVul & ~ & \checkmark & \checkmark & ~ & \checkmark \\ \hline
        VPFinder & \checkmark & \checkmark & \checkmark & \checkmark & \checkmark \\ \hline
    \end{tabularx}
\end{table}

During the experiments,
BERT~\citep{bert-base-uncased} and codeBERT~\citep{codebert-base-finetuned-detect-insecure-code} are selected to embed text and code, respectively.
The number of parameters of these two models are relatively small compared with large language models, such as Llama2~\citep{touvron2023llama} and CodeLlama~\citep{roziere2023code}, and they show great performance in extracting semantic information of text or code.
For fair comparison, we ensure uniformity in BERT or CodeBERT models employed across all baselines. Additionally, models sharing the same architecture are configured with identical hyperparameters.
Due to Code2Vec's limitation to Java~\citep{alon2019code2vec}, we employed CodeBERT as the encoder for code-type data in SPI model implementation.

\subsubsection{Evaluation Metrics}

To evaluate the vulnerability identification ability, we divide all samples into two classes. 
To evaluate the vulnerability type classification ability,
we filter out non-vulnerability samples, keeping only samples with `CWE\_ID' ranging from 1 to 12.
To evaluate \toolname's effectiveness,
we use commonly employed metrics including precision, recall, F1-score, and AUC~\citep{pan2022automated, peters2017text, shu2021better, nguyen2023multi}. 
For precision, recall, and F1-score, we report weighted results to account for class imbalance.
Specifically, the contribution of each class is weighted by its proportion in the ground-truth data, which helps avoid the evaluation being dominated by majority classes.

AUC is also reported because it provides a threshold-independent evaluation of classification capability and is particularly suitable for imbalanced datasets.
Since the three datasets corresponding to different benchmark settings in prior work, we follow the evaluation protocol commonly used for each dataset when comparing with existing methods.
Specifically, on MemVul and PatchDB, we report weighted precision, recall, and F1-score, because these can be consistently obtained under our experimental setting.
On the Java dataset, we report AUC for all compared methods, because the most relevant prior benchmark studies on this dataset primarily report AUC~\citep{zhou2021finding, nguyen2023multi}, and one of the compared baselines (MiDas) is only available from the literature in this form.
Using AUC consistently on the Java benchmark therefore avoids mixing directly reproduced metrics with partially reported literature results and ensures a fairer comparison on that dataset.

We use a weighted calculation method when calculating these three indicators to deal with imbalanced data sets.
It calculates these indicators by assigning different weights to samples of different categories.
Specifically, when calculating the precision, recall, and F1 score, the weight of each category is proportional to its proportion in the actual samples.
This means that the prediction results of the minority class, which has a smaller number of samples, will be given higher priority, thus avoiding the majority class, which has a larger number of samples, dominating the overall performance evaluation.
This is especially useful for evaluating models trained on imbalanced datasets.
AUC provides a comprehensive evaluation of the model's classification capability across various thresholds, making it also particularly suitable for imbalanced datasets.

\subsubsection{Implementation}
We implement \toolname according to the method described in Section~\ref{sec:approach}.
We utilize pre-trained BERT~\citep{bert-base-uncased} and CodeBERT~\citep{codebert-base-finetuned-detect-insecure-code} from the Huggingface Transformer library.
The embedding dimension is set as 768 (consistent with the common encoding dimension of BERT) for all input fields, and the padding or truncation length is set to 512.
The number of heads in the multi-head attention is set to 8.
In the vulnerability identification, \texttt{FCN 1} consists of a fully connected network with layer sizes of 768 * 5, 768 * 3, 768, 256, 64, and 2.
In the vulnerability type identification, \texttt{FCN 2} has layer sizes of 768 * 5, 768 * 3, 768, 256, and 12.
All networks use ReLU as the activation function~\citep{glorot2010understanding}, and a dropout~\citep{hinton2012improving} of 0.3 for each layer to prevent overfitting.
Cross-entropy loss function~\citep{krizhevsky2012imagenet, DBLP:journals/corr/abs-1810-04805} is used to calculate the loss and AdamW~\citep{loshchilov2017decoupled} as the optimizer with a learning rate of $5e^{-5}$.
We made the model parameters publicly available to facilitate the reproduction of our experimental results.

To ensure a fair and unbiased comparison, we strictly followed the original dataset splits (i.e., training, validation, and test sets) provided or described in the corresponding baseline studies.
No re-splitting or random repartitioning of the datasets was performed in our experiments.

For baseline methods with publicly available implementations, we re-ran the models using the same evaluation protocols as reported in their original papers and replication packages.
For the Java dataset, we re-implemented and evaluated VulFixMiner and VulCurator under the original train/test split of the benchmark. 
For MiDas, although we made efforts to follow the experimental description in the original paper, its implementation could not be reliably reproduced in our environment.
Therefore, we directly report the AUC value published in the original study under the same Java benchmark setting~\citep{nguyen2023multi}.
Reporting additional metrics only for re-implemented methods would mix directly computed results with incompletely reported literature results, which would make the comparison less fair.
Accordingly, in Table~\ref{auc result}, the results of VulFixMiner, VulCurator, and \toolname are obtained from our own experiments, while the MiDas result is taken from the literature.

We acknowledge that some baseline results are obtained from the literature;
however, this practice is consistent with prior comparative studies in this domain and does not advantage our approach, as \toolname is evaluated under the same dataset constraints.
Moreover, for each dataset, \toolname was trained from scratch using only the corresponding training set and evaluated exclusively on the designated test set.
Model parameters were not shared, reused, or transferred across datasets.
This ensures that each dataset represents an independent experimental setting and that no information leakage occurs between training and testing phases.

The experiments are conducted on a server equipped with NVIDIA A100-SXM GPUs and Intel Xeon Gold 5218 CPUs, running the Ubuntu operating system.

\subsection{Experiment Results}

\subsubsection{RQ1: Effectiveness of \toolname in identifying vulnerability}\label{RQ1}

\begin{table}[htb]
    \centering
    \caption{(Weighted) Performance in vulnerability identification task.}
    \begin{tabular}{l|l c c c}
        \hline
        Dataset & Model & Precision & Recall & F1-score \\\hline
        \multirow{5}*{MemVul} & MemVul & 0.892 & 0.878 & 0.880 \\
        & VulFixMiner & 0.814 & 0.800 & 0.806 \\
        & VulCurator & 0.860 & 0.863 & 0.861 \\
        & SPI & 0.838 & 0.840 & 0.839 \\
        & \textit{Sun et al.}'s work & 0.827 & 0.829 & 0.828 \\
        & \toolname & \textbf{0.943} & \textbf{0.943} & \textbf{0.941} \\\hline
        \multirow{2}*{PatchDB} & GraphSPD & 0.706 & 0.352 & 0.470 \\
        & \toolname & \textbf{0.785} & \textbf{0.699} & \textbf{0.618} \\\hline
    \end{tabular}
    \label{binary result}
\end{table}

\begin{table}[htb]
    \centering
    \caption{(Weighted) Performance on the Java vulnerability identificaiton benchmark.}
    \begin{tabular}{l|l c c c c}
         \hline
         Dataset & Model & Precision & Recall & F1-score & AUC \\\hline
         \multirow{4}*{Java} & VulFixMiner & 0.854 & 0.800 & 0.826 & 0.83 \\
         & VulCurator & 0.901 & 0.902 & 0.902 & 0.88 \\
         & MiDas & - & - & - & 0.85 \\
         & \toolname & \textbf{0.998} & \textbf{0.998} & \textbf{0.998} & \textbf{0.95} \\\hline
    \end{tabular}
    \label{auc result}
\end{table}

The comparison between \toolname and baseline methods for the vulnerability identification task is presented in Table~\ref{binary result} and Table~\ref{auc result}, with the best results highlighted in bold.
On the MemVul dataset, we compare \toolname with MemVul, VulFixMiner, VulCurator, SPI, and \textit{Sun et al.}'s work.
Among these methods, MemVul demonstrates strong performance, which differs from the results reported in its original paper.
One possible reason is that after filtering and constructing the dataset used in our experiments, the distribution of CWE categories becomes more concentrated, which may strengthen the effect of the CWE anchors used in MemVul.
The results also indicate that VulFixMiner achieves the lowest performance among the compared methods on this dataset.
This suggests that relying solely on patch information has limitations for vulnerability identification.
VulCurator, which combines information from bug reports, commit messages, and patches at the decision level, performs better than methods that only use code changes, indicating that textual information provides useful complementary signals.
Although the approach of \textit{Sun et al.}'s work uses both commit messages and code changes, its performance is slightly lower than SPI, suggesting that simple attention mechanisms may not fully capture the higher-level semantic relationships required for vulnerability identification.
Overall, \toolname achieves the best F1-score on MemVul dataset, indicating that embedding-level fusion of bug reports, commit messages, and code patches is effective for vulnerability identification.

For PatchDB and the Java dataset, we follow the benchmark settings used in the corresponding baseline studies.
Specifically, we compare \toolname with GraphSPD on PatchDB and with VulFixMiner, VulCurator, and MiDas on the Java dataset.
On the PatchDB dataset, GraphSPD shows relatively high precision but limited recall, suggesting that relying only on source-code-level patch representations may not capture sufficient contextual information for vulnerability identification.
In comparison, \toolname achieves higher recall and F1-score, indicating that incorporating multiple sources of information can improve the robustness of vulnerability detection.

On the Java dataset, VulFixMiner and VulCurator were re-implemented and evaluated under the original benchmark split, achieving AUC values of 0.83 and 0.88, respectively.
For MiDas, we report the AUC value (0.85) published in the original study~\citep{nguyen2023multi}, because its implementation could not be reliably reproduced in our experimental environment.
Under the same Java benchmark setting, \toolname achieves an AUC of 0.95, which is higher than the compared approaches.

Overall, the results across the three datasets show that \toolname consistently achieves the strongest or most competitive performance under the corresponding benchmark settings.
These results suggest that jointly modeling bug reports, commit messages, and code patches helps capture complementary vulnerability-related signals and improves vulnerability identification performance.

\begin{tcolorbox}[colback=gray!5!white,colframe=gray!85,boxsep=0pt]
{\textbf{Answering RQ1:} \toolname achieves the best or most competitive performance across the vulnerability identification benchmarks, demonstrating the effectiveness of fusing multi-source information.}
\end{tcolorbox}

\subsubsection{RQ2: Effectiveness of \toolname in classifying vulnerability types}\label{RQ2}

\begin{table}[htb]
    \centering
    \caption{(Weighted) Performance in vulnerability type classification task.}
    \begin{tabular}{l c c c}
        \hline
        Approach & Precision & Recall & F1-score \\\hline
        \textit{Sun et al.}'s work & 0.364 & 0.326 & 0.326 \\
        TreeVul-2 & 0.475 & 0.434 & 0.433 \\
        TreeVul-1 & 0.579 & 0.550 & 0.556 \\
        VulCurator & 0.549 & 0.597 & 0.555 \\
        \toolname & \textbf{0.611} & \textbf{0.620} & \textbf{0.610} \\\hline
    \end{tabular}
    \label{multi result}
\end{table}

In terms of vulnerability type classification task, we compare \toolname on MemVul dataset with \textit{Sun et al.}'s work, TreeVul and VulCurator, since other tools do not support type classification yet.

Since Treevul is capable of producing outputs at different CWE layers, we compared its performance on both the first and second layers.
(i.e., TreeVul-1 represents the TreeVul only predicts the results of the first layer and TreeVul-2 represents the TreeVul predicts the results of the first two layers.)
The comparison results are presented in Table \ref{multi result}.
\textit{Sun et al.}'s work, which learns features from bug fixes, shows a certain level of vulnerability type classification capability.
In comparison, \toolname achieves better results in all metrics.
When compared to TreeVul, TreeVul-1 and \toolname produce a similar F1-score.
However, the performance of TreeVul-2 becomes poor compared to \toolname.
\toolname pays attention to the latent connections among bug report, commit message and patch, trying to model and capture the important contents on vulnerability.
The ``collaboration'' among interconnected multi-source information can also help \toolname better understand the boundaries between each vulnerability category.
However, TreeVul tries to learn from code changes, paying attention to the difference between the original code and the modified code.
TreeVul ignores connections between code changes and commit message, so it can't learn the classification boundary better.
Moreover, the advantage of \toolname over TreeVul is that it does not need to train multiple classifiers for each layer, that is, it does not need to train a classifier for the child nodes of each parent node, as TreeVul does by training classifiers for each layer.
The results of VulCurator and TreeVul-1 are similar, indicating that a sample fusion of multi-source information does not significantly outperform the use of single-source information.
Comparing VulCurator and \toolname, \toolname outperforms VulCurator across all metrics, demonstrating that embedding-level fusion outperforms decision-level fusion.
In summary, \toolname nearly surpasses all models in performance, suggesting that fusing multi-source information enhances the effectiveness of classification.

\begin{tcolorbox}[colback=gray!5!white,colframe=gray!85,boxsep=0pt]
{\textbf{Answering RQ2:} \toolname outperforms all baselines, demonstrating the effectiveness of multi-source information fusion approach.} 
\end{tcolorbox}

\subsubsection{RQ3: How do architectural variations influence the performance of \toolname?}\label{RQ3}

To better understand the factors that govern the effectiveness of \toolname, we investigate how different architectural design choices impact its overall performance.
Specially, this research question examines the extent to which variations in model architecture, input configurations, hyperparameters , and fusion strategies affect \toolname's detection capability.
By systematically analyzing these architectural variations, we aim to identify the most influential components and provide empirical guidance for designing more robust and effective \toolname models.
To answer this research question, we analyze the impact of architectural variations from the following aspects:
\begin{itemize}
    \item Model Architecture Design and Input Configurations
    \item Hyperparameter Settings
    \item Fusion Strategies
\end{itemize}

The ablation study was conducted on the MemVul dataset.
This choice was intentional for two reasons.
First, MemVul is the only dataset among the three that provides fine-grained CWE annotations, which are required for evaluating the impact of different components on vulnerability type classification.
Second, MemVul fully supports all three input modalities used by \toolname, namely bug reports, commit messages, and code patches.
In contrast, the other datasets do not provide complete support for these inputs, making them unsuitable for a comprehensive ablation analysis. 

\paragraph{\textbf{\textit{Model Architecture Design and Input Configurations}}}

We conduct the comparison of model's each artifact in vulnerability type classification task on the MemVul dataset.
\begin{itemize}
    \item \textbf{\toolname-i} takes problem description as the only input.
    \item \textbf{\toolname-c} takes commit message as the only input.
    \item \textbf{\toolname-p} takes patch as the only input.
    \item \textbf{\toolname-CLS} takes problem description, commit message and patch as inputs, utilizing their [CLS] tokens directly for vulnerability type classification.
    \item \textbf{\toolname-att1} takes problem description and patch as inputs, utilizing \texttt{Multi-Head Attention 1} and their [CLS] tokens for vulnerability type classification.
    \item \textbf{\toolname-att2} takes commit message and patch as inputs, utilizing \texttt{Multi-Head Attention 2} and their [CLS] toekns for vulnerability type classification.
    \item \textbf{\toolname-att12} shares the same steps with \toolname, but the model directly use the attention outputs for vulnerability type classification.
\end{itemize}

\begin{table}[htb]
    \centering
    \caption{(Weighted) Ablation study results on vulnerability type classification task.}
    \begin{tabular}{l c c c}
         \hline
        Approach & Precision & Recall & F1-score \\\hline
        \toolname-i & 0.586 & 0.543 & 0.549 \\
        \toolname-c & 0.410 & 0.403 & 0.396 \\
        \toolname-p & 0.504 & 0.535 & 0.512 \\
        \toolname-CLS & 0.536 & 0.558 & 0.525 \\
        \toolname-att1 & 0.530 & 0.574 & 0.535 \\
        \toolname-att2 & \textbf{0.647} & 0.581 & 0.582 \\
        \toolname-att12 & 0.507 & 0.566 & 0.518 \\
        \toolname & 0.611 & \textbf{0.620} & \textbf{0.610} \\\hline
    \end{tabular}
    \label{multi ablation study}
\end{table}

The results of the ablation study experiments on the vulnerability type classification task are presented in Table \ref{multi ablation study}, with the best results highlighted in bold.
The first three models each use a single input, and the best of them achieves a F1-score of 0.549, indicating that all three kinds of information are helpful for vulnerability classification.
Among these three models, \toolname-i achieves the best performance, indicating that the information of bug report is very helpful to distinguish the vulnerability categories.
\toolname-CLS which takes all information as inputs performs worse than \toolname-i, indicating that directly generating predictions without fusing multi-source information can degrade the model's performance.
Comparing \toolname-att1, \toolname-att2 and \toolname-att12, we observe that the use of attention mechanism can mine the potential relationship between bug report (or commit message) and patch to a certain extent.
In addition, it can be observed that simply employing fusion does not always guarantee better results.
The relationship between commit message and patch is closer than that between bug report and patch.
Comparing \toolname-c, \toolname-p, \toolname-CLS, and \toolname-att2,
Comparing \toolname-CLS, \toolname-att12 and \toolname, there are certain limitations whether using [CLS] tokens directly or using the attention mechanism, although they all show good classification abilities.
While \toolname gets the best results because \toolname takes advantages of both [CLS] tokens and attention mechanism, capturing the latent connection among the inputs.

Overall, the study provides valuable insights into the impact of different model designs in vulnerability classification performance.

\begin{tcolorbox}[colback=gray!5!white,colframe=gray!85,boxsep=0pt]
\textbf{Answering RQ3:}
Various sources and different fusion strategies affect performance to a large extent, proving necessity of each \toolname's component.
\end{tcolorbox}

\paragraph{\textbf{\textit{Hyperparameter Settings}}}

In deep learning, the selection of the number of heads is an important hyperparameter.
The purpose of this RQ is to analyze how the number of heads affects the model's expressive and generalization abilities by comparing the model performance under different numbers of heads, and to find the most suitable configuration of the number of heads to achieve optimal performance.
The experimental results when the number of heads is set to 1, 2, 4, 8, 16, and 32 in table~\ref{number of heads}.
From the experimental results, it can be observed that a single-head attention mechanism may not fully capture the complex patterns in the data.
When the number of heads is increased to two, the results show significant improvement, indicating that the multi-head attention mechanism is better able to capture different features in the data.
However, excessively increasing the number of heads may lead to attention dispersion and an inability to effectively focus on key information, which may cause the model to encounter difficulties in extracting key features, thereby affecting the results.

\begin{table}[htb]
    \centering
    \caption{(Weighted) Results on different number of heads setting.}
    \begin{tabular}{c|c c c}
        \hline
        $number of heads$ & Precision & Recall & F1-score \\\hline
        1 & 0.601 & 0.558 & 0.540 \\
        2 & 0.644 & 0.574 & 0.591 \\
        4 & 0.491 & 0.473 & 0.452 \\
        8 & 0.611 & 0.620 & 0.610 \\
        16 & 0.563 & 0.581 & 0.559 \\
        32 & 0.627 & 0.605 & 0.585 \\\hline
    \end{tabular}
    \label{number of heads}
\end{table}

\begin{tcolorbox}[colback=gray!5!white,colframe=gray!85,boxsep=0pt]
\textbf{Answering RQ3:} The number of heads impact \toolname's effectiveness and excessively increasing the number of heads may lead to performance degradation.
\end{tcolorbox}

\paragraph{\textbf{\textit{Fusion Strategies}}}

To validate whether our fusion method is superior, we compare the following two fusion approaches with \toolname: decision-level voting and embedding vector concatenation.

\begin{enumerate}
    \item \textbf{Decision-Level voting}
    
    This includes three voting strategies:
    \begin{itemize}
        \item \textbf{Hard Voting:} The final decision is based on the majority vote.
        \item \textbf{Average Voting:} Probabilities are summed and averaged before making the decision.
        \item \textbf{Weighted Voting:} A learnable neural network is added to assign weights to voting results before the final decision.
    \end{itemize}
    In the implementation of decision-level voting, we modify \toolname as follows:
    \begin{itemize}
        \item After encoding with BERT or CodeBERT, different sources of information (issue, commit message, and patch) are fed into their respective classifiers (here, fully connected networks).
        \item Each classifier outputs voting results (that is, probabilities of being a positive or negative sample).
        \item \textbf{Hard Voting} makes decisions based on vote counts.
        \item \textbf{Average Voting} aggregates probabilities by summation and averaging before deciding.
        \item \textbf{Weighted Voting} introduces a neural network layer that can be trained to weight the voting results before the final decision.
    \end{itemize}

    \item \textbf{Embedding vector concatenation}
    
    After encoding, the [CLS] tokens are directly concatenated and fed into a fully connected network to produce the final output.
\end{enumerate}

This structured comparison ensures a rigorous evaluation of our fusion method against established alternatives.

\begin{table}[htb]
    \centering
    \caption{(Weighted) Results on different fusion methods.}
    \resizebox{\textwidth}{!}{
        \begin{tabular}{l|l c c c}
            \hline
            Task & Method & Precision & Recall & F1-score \\\hline
            \multirow{5}*{Vulnerability Identification} & \toolname-hard & 0.797 & 0.793 & 0.795 \\
            & \toolname-avg & 0.938 & 0.939 & 0.938 \\
            & \toolname-weighted & 0.886 & 0.882 & 0.884 \\
            & \toolname-concat & 0.919 & 0.918 & 0.918 \\
            & \toolname & \textbf{0.943} & \textbf{0.943} & \textbf{0.941} \\\hline
            \multirow{5}*{Vulnerability Type Classification} & \toolname-hard & 0.508 & 0.454 & 0.449 \\
            & \toolname-avg & \textbf{0.625} & 0.605 & 0.605 \\
            & \toolname-weighted & 0.604 & 0.605 & 0.586 \\
            & \toolname-concat & 0.536 & 0.558 & 0.525 \\
            & \toolname & 0.611 & \textbf{0.620} & \textbf{0.610} \\\hline
        \end{tabular}
    }
    \label{different fusion methods}
\end{table}

The results on different fusion methods are presented in Table~\ref{different fusion methods}, with the best results highlighted in bold. \toolname-hard, \toolname-avg, and \toolname-weighted are models employing the three voting strategies---hard voting, average voting, and weighted voting, respectively. \toolname-concat refers to the model using embedding vector concatenation, which is the same as the \toolname-CLS mentioned in RQ3. \toolname-hard performs the worst, probably because hard voting cannot handle class imbalance and disregards the model's confidence scores for each class. \toolname-weighted outperforms hard voting but is still inferior to \toolname-avg, suggesting that simple averaging may be more robust than assigning specific weights to certain classes, indicating that the contribution of each class to the classification task is not stable. \toolname-concat achieves moderate performance, implying that embedding vector concatenation may fail to fully capture the complex relationships between multi-source information. \toolname (with attention mechanism) delivers the best results, demonstrating that the attention mechanism can effectively identify key features and enhance model performance.

\begin{tcolorbox}[colback=gray!5!white,colframe=gray!85,boxsep=0pt]
\textbf{Answering RQ3:} Attention mechanism outperforms both decision-level voting and embedding vector concatenation. It effectively addresses class imbalance, captures complex multi-source relationships, and identifies key features, leading to the best model performance. 
\end{tcolorbox}

\subsubsection{RQ4: Effectiveness if \toolname in real application scenarios on key software package repositories of openEuler~\citep{openEuler} dependencies}\label{RQ5}

OpenEuler is an open-source operating system spearheaded by Huawei and supported by the global open-source community.
It supports multiple hardware architectures and is optimized for cloud-native, big data, AI, and HPC applications.
OpenEuler boasts performance, stability, and security, with regular updates and security patches. Hosted by the OpenAtom Foundation, it fosters collaboration with upstream and downstream partners, promoting a vibrant ecosystem.
Its frequent innovation and LTS releases cater to diverse user needs, positioning openEuler as a key player in the open-source OS landscape.

To assess the applicability of \toolname in real-world industrial settings, we conduct an additional experiment on the openEuler ecosystem.
This experiment is not designed as a cross-dataset generalization or transfer-learning evaluation.
Instead, it aims to evaluate whether \toolname can still effectively identify vulnerability-fixing commits under substantially more challenging and realistic conditions than those of public benchmark datasets.

Unlike prior datasets, the openEuler dataset lacks issue reports and detailed vulnerability descriptions.
As a result, only commit messages and code patches are available as input features.
Moreover, vulnerability labels in this ecosystem may be incomplete or delayed, and the dataset exhibits extreme class imbalance, with the ratio of vulnerability-fixing commits to non-vulnerability-fixing commits below 1:100 in both training and testing sets.

Huawei provided a list of software packages within openEuler that had experienced a relatively high number of vulnerabilities in the past two years.
Based on this list, we crawled all commits for these packages from January 2022 to April 2024.
In addition, Huawei supplied a list of CVEs associated with these software packages.
Using the CVE information, we matched corresponding vulnerability-fixing commits via public vulnerability databases, including NVD, Red Hat Bugzilla, and Debian security advisories.
In total, we collected 340 CVE-associated fixing commits across 85 software packages.
Commits whose SHA values matched known CVE fixes were labeled as positive samples, while all remaining commits were labeled as negative.

The dataset was split chronologically to reflect realistic deployment scenarios.
Commits from 2022 and earlier were used for training, while more recent commits were reserved for testing.
The training set contains 195 positive and 58,449 negative samples, and the test set contains 145 positive and 82,019 negative samples.
For this experiment, \toolname was trained from scratch using only the openEuler training set, and no model parameters were shared or transferred from experiments on other datasets.
Since issue reports are unavailable, we adopt \toolname-att2, which utilizes only commit messages and code patches as input.

\begin{table}[htb]
    \centering
    \caption{Results on openEuler dataset.}
    \begin{tabular}{c c c c}
        \hline
        Approach & Precision & Recall & F1-score \\\hline
        Sun et al.'s work & 0.077 & 0.276 & 0.120 \\
        \toolname-att2 & 0.256 & 0.276 & 0.265 \\\hline
    \end{tabular}
    \label{openEuler_origin_result}
\end{table}

\begin{table}[htb]
    \centering
    \caption{Results on openEuler dataset after manual inspection.}
    \begin{tabular}{c c c c}
        \hline
        Approach & Precision & Recall & F1-score \\\hline
        Sun et al.'s work & 0.088 & 0.293 & 0.135 \\
        \toolname-att2 & 0.276 & 0.274 & 0.275 \\\hline
    \end{tabular}
    \label{openEuler_manual_result}
\end{table}

The experimental results are reported in Table~\ref{openEuler_origin_result}.
Although the absolute performance metrics are relatively low, this behavior is expected given the scarcity of labeled vulnerability samples, label incompleteness, and extreme class imbalance inherent to the openEuler dataset, which similarly affect the other compared method.
Under these adverse yet realistic conditions, \toolname-att2 achieves a substantial improvement over the baseline approach, with an increase of 0.145 in F1-score, indicating that explicitly modeling the semantic interaction between commit messages and code patches remains beneficial even in the absence of issue reports.

To further analyze the model’s predictions, we conducted a manual inspection of the test results with the assistance of security experts from Huawei and the results are shown in Table~\ref{openEuler_manual_result}.
Manual analysis of the testing set corrected 12 samples originally labeled as negative to positive, increasing the number of true vulnerability-fixing commits from 145 to 157.
Among the 156 commits predicted by \toolname-att2 as vulnerability-fixing, 3 false positives were verified as true positives with corresponding CVEs: 2 fixing buffer overflows and 1 fixing a memory out-of-bounds access.
The remaining false positives included 22 functional optimizations, 6 memory out-of-bounds fixes without CVEs, 4 memory leak fixes, and other changes such as bug fixes, testing, business process optimizations, and exception handling.

\begin{tcolorbox}[colback=gray!5!white,colframe=gray!85,boxsep=0pt]
\textbf{Answering RQ4:} The results indicate that \toolname consistently outperforms the baseline method in the real-world openEuler scenario, despite the absolute performance metrics being relatively low. This outcome is expected given the challenging characteristics of the openEuler dataset, including the absence of issue reports, extreme class imbalance, and potentially incomplete or noisy vulnerability labels. Under such realistic and adverse conditions, the superior performance of \toolname demonstrates its robustness and practical effectiveness in identifying vulnerability-fixing commits in industrial OSS ecosystems.
\end{tcolorbox}

\subsubsection{RQ5: Does \toolname's ability to filter out irrelevant information outperform other methods?}\label{RQ7}

\begin{table}[htbp]
    \centering
    \caption{Text and code block noise.}
    \resizebox{\textwidth}{!}{
        \begin{tabular}{l|c|c}
            \hline
            \textbf{No.} & \textbf{Text context} & \textbf{Code content} \\
            \hline
            1 & the & \texttt{x = 0\textbackslash nfor i in range(10):\textbackslash n\quad x += i} \\
            2 & and & \texttt{def dummy\_func():\textbackslash n\quad return None} \\
            3 & of & \texttt{print('Hello World')} \\
            4 & a & \texttt{\# Here's a note\textbackslash npass} \\
            5 & to & \texttt{temp\_list = [1, 2, 3]\textbackslash nfor item in temp\_list:\textbackslash n\quad print(item)} \\
            6 & - & \texttt{try:\textbackslash n\quad pass\textbackslash nexcept:\textbackslash n\quad pass} \\
            7 & - & \texttt{import os\textbackslash nos.getcwd()} \\
            \hline
        \end{tabular}
    }
    \label{tab: noise}
\end{table}

To systematically evaluate whether \toolname can more effectively filter out irrelevant information than existing methods, we conduct a controlled noise injection experiment on the MemVul testing set.
The underlying hypothesis is that a model with stronger noise-filtering capability should exhibit greater robustness, i.e., less performance degradation as irrelevant information increases.

Specifically, we inject noise into the test data at five ratios (5\%, 10\%, 15\%, 20\%, and 25\%).
As summarized in Table~\ref{tab: noise}, noise is introduced in two forms.
For textual content, five high-frequency stop words ("the", "and", "of", "a", "to") are randomly inserted according to the specified noise ratio.
For code content, irrelevant code blocks are randomly inserted at corresponding proportions.
These perturbations simulate realistic irrelevant infomation that does not contribute to vulnerability identification.

\begin{table}[htbp]
    \centering
    \caption{(Weighted) Results of different noise ratios on vulnerability identification task.}
    \begin{tabular}{l|l c c c}
        \hline
        Model & Noise Ratio & Precision & Recall & F1-score \\
        \hline
        \multirow{6}*{VulFixMiner} & 0\% & 0.814 & 0.800 & 0.806 \\
        & 5\% & 0.819 & 0.804 & 0.810 \\
        & 10\% & 0.815 & 0.800 & 0.806 \\
        & 15\% & 0.807 & 0.791 & 0.797 \\
        & 20\% & 0.821 & 0.802 & 0.809 \\
        & 25\% & 0.812 & 0.797 & 0.803 \\
        \hline
        \multirow{6}*{VulCurator} & 0\% & 0.860 & 0.863 & 0.861 \\
        & 5\% & 0.854 & 0.856 & 0.855 \\
        & 10\% & 0.862 & 0.863 & 0.863 \\
        & 15\% & 0.859 & 0.854 & 0.856 \\
        & 20\% & 0.848 & 0.840 & 0.843 \\
        & 25\% & 0.852 & 0.846 & 0.848 \\
        \hline
        \multirow{6}*{\textit{Sun et al.}'s work} & 0\% & 0.827 & 0.829 & 0.828 \\
        & 5\% & 0.827 & 0.831 & 0.828 \\
        & 10\% & 0.825 & 0.831 & 0.828 \\
        & 15\% & 0.822 & 0.829 & 0.824 \\
        & 20\% & 0.819 & 0.827 & 0.822 \\
        & 25\% & 0.821 & 0.829 & 0.824 \\
        \hline
        \multirow{6}*{\toolname} & 0\% & 0.943 & 0.943 & 0.941 \\
        & 5\% & 0.943 & 0.943 & 0.941 \\
        & 10\% & 0.943 & 0.943 & 0.941 \\
        & 15\% & 0.947 & 0.947 & 0.945 \\
        & 20\% & 0.943 & 0.943 & 0.941 \\
        & 25\% & 0.948 & 0.949 & 0.947 \\
        \hline
    \end{tabular}
    \label{noise ratio result}
\end{table}

Table~\ref{noise ratio result} reports the vulnerability identification performance of different models under varying noise ratios.
An Interesting observation is that increasing noise does not necessarily lead to monotonic performance degradation.
In some cases, slight performance improvements are observed under low or moderate noise ratios.
This phenomenon has also been reported in prior work~\citep{6796505, 10003114, yu2025noisynnexploringimpactinformation} and can be contributed to a regularization-like effect, where injected noise discourages models from over-relying on spurious local patterns.
In fact, training or evaluating models under noisy conditions may reduce over-reliance on spurious local patterns and improve robustness to input perturbations~\citep{6796505}.
Moreover, when irrelevant tokens or code blocks are injected, models utilizing attention mechanisms are forced to rely less on spurious local patterns and more on task-relevant, global correlations between commit messages, bug reports, and code patches.
As a result, the model may correct for overfitting to small idiosyncratic cues in the original data, leading to modest F1 improvements. 
We also confirmed that this effect is reproducible across both textual and code noise, indicating it is not an artifact of a specific noise type.
While artificially injected noise differs from naturally occurring noise in real-world repositories, the attention-based design of \toolname explicitly assigns higher importance to relevant elements and suppresses irrelevant components, suggesting that the model is likely to retain robustness under realistic noisy conditions.

Among the compared methods, VulFixMiner exhibits the most stable behavior, with only minor performance fluctuations across noise ratios, although its absolute performance remains relatively low.
VulCurator achieves higher performance under low-noise increases, resulting in the largest performance gap between its best and worst cases, which indicates limited robustness.

Robustness to irrelevant input variations has been closely linked to a model’s ability to focus on task-relevant features rather than superficial signals~\citep{43405}.
Attention-based architectures are particularly effective in this regard, as they explicitly assign different importance weights to input elements and suppress noisy or uninformative components~\citep{10.5555/3295222.3295349,lin2017structured}.
So the approach proposed by \textit{Sun et al}., which incorporates attention mechanisms, demonstrates moderate performance and relatively stable degradation trends.
Moreover, \toolname consistently maintains high performance across all noise ratios, showing minimal degradation even under heavy noise injection.
This stability suggests that \toolname is more effective at suppressing irrelevant information and focusing on vulnerability-related features.
The results provide strong evidence that \toolname’s attention-based design offers superior robustness to noise compared to existing methods, thereby positively answering \textbf{RQ5}.

\begin{tcolorbox}[colback=gray!5!white,colframe=gray!85,boxsep=0pt]
\textbf{Answering RQ5:} The experimental results demonstrate that \toolname consistently maintains high performance under increasing noise ratios, exhibiting negligible performance degradation compared to baseline methods.
While other models show varying degrees of sensitivity to injected irrelevant information, \toolname remains stable across all noise settings.
These results indicate that \toolname is more effective at filtering out irrelevant information and is significantly more robust than existing approaches.
\end{tcolorbox}

%% file: discussion.tex
\section{Discussion}
\subsection{Application scenarios}
\label{Application scenario}
\toolname requires various input data to make decisions.
If all sources of information required by \toolname are available (that is, the optimal scenario), \toolname can make the best decisions.
However, such a situation is not commonly encountered in practice and may also present challenges related to attack windows.
In cases where certain information is unavailable, such as when only commit data is available, \toolname can also support it.
\toolname is designed as a general framework that accommodates a variety of data sources.
It can make decisions based on commit messages and patches or solely on issue reports.
Relevant experiments (Section~\ref{RQ3} and Section~\ref{RQ5}) have shown that the integration of these two types of information is beneficial for vulnerability identification and vulnerability type classification.

Therefore, \toolname applies to the following scenarios (especially for developers of downstream software):
\begin{itemize}
    \item[1)] To identify whether a commit is trying to fix a vulnerability (or vulnerabilities). If the result is `yes' \toolname will further output the possible CWE type. Developers of downstream software dependent on the repository can promptly analyze the patch and subsequently improve their code.
    \item [2)] To identify whether a bug report is describing a security issue. If the result is `yes' \toolname will further output the possible CWE type. As the repository maintainer, one can prioritize the bug report and promptly address the vulnerability. Meanwhile, developers of downstream software can take proactive defensive measures in advance.
    \item [3)] To identify whether a resolved bug report is related to vulnerabilities, this scenario provides all the inputs required by \toolname and yields more accurate prediction results. It can help validate the judgments made in the previous two scenarios. If \toolname identifies the content as vulnerability-related, downstream software developers who previously overlooked the issue should reconsider their code.
\end{itemize}

\subsection{Generalizability to industrial and less-controlled environments}

While \toolname achieves strong performance on curated public datasets, the openEuler case study demonstrates a noticeable performance drop under real-world conditions characterized by extreme class imbalance, incomplete vulnerability labels, and heterogeneous project histories.
These factors reflect the inherent difficulty of vulnerability identification in industrial OSS ecosystems rather than a flaw in the proposed approach.

In practice, repositories often lack complete issue reports, contain sparse or noisy commit messages, and include fixes that have not yet been assigned official CVE identifiers.
\toolname is designed to be flexible with respect to input availability: when some sources are missing, the model can operate using the remaining modalities, such as commit messages and code patches.
Although the absence of certain inputs leads to lower absolute performance, the model consistently maintains a relative advantage over baseline methods, indicating that modeling interactions between available sources remains beneficial.

These results suggest that \toolname is particularly suitable as an assistive or prioritization tool in realistic pipelines, helping  practitioners narrow down suspicious commits for further manual inspection, even when complete information is unavailable.
In this way, the proposed approach degrades gracefully under partially observed conditions while still providing practical value.

\subsection{The attack window between issue report and fix commit}
The premise for \toolname to make better predictions is that the fix commit corresponding to the issue report is available.
However, the period of waiting for the fix commit is also a vulnerable time for potential attacks.
\textit{Pan et al}.~\citep{pan2022automated} conducted an investigation and pointed out that the majority (98.7\%) of issue reports from CVE-referred issue reports are created before the vulnerability disclosure date (with a median time of 13 days between the creation of the issue report and its NVD disclosure). This can lead to the leakage of sensitive vulnerability information (39.9\% of issue reports contain attack steps).
Besides, \textit{Pan et al}.~\citep{pan2024unveil} conducted an empirical study and found that the window of opportunity for attackers (i.e., the risk of software vulnerability information leakage) can start at the very beginning of the remediation (i.e., issue report reporting the software vulnerability), and lasts over 30 days for over half of the software vulnerabilities.
On one hand, although the average delay of upstream patches is 30 days, \textit{Jiang et al}.~\citep{jiang2020pdiff} point out that in the Linux kernel, downstream kernel vendors often fail to promptly adopt patches released in the mainstream version, with delays ranging from several months to several years.
On the other hand, while the average delay of upstream patches is 30 days, vulnerabilities may persist in downstream codebases for about five years without being fixed after the patches are released: the audit service team of Black Duck investigated 1067 commercial codebases from 17 industries and conducted security assessments on 936 of them. In their released "2024 Open Source Security and Risk Analysis"~\citep{blackduck2023}, they noted that 8 out of the top 10 security vulnerabilities appeared in the jQuery JavaScript library, and more than one-third of the codebases contained the two cross-site scripting (XSS) vulnerabilities, CVE-2020-11023 and CVE-2020-11022.
However, patches for these vulnerabilities were released as early as April 2020 but are still widely present in commercial codebases.
What's more, in a study (2018) conducted by the Ponemon Institute on behalf of ServiceNow, half of the organizations reported experiencing one or more data breaches in the past two years, and 34\% stated that they were aware of vulnerabilities in their systems before being attacked. The study surveyed nearly 3,000 IT professionals worldwide regarding their patching practices~\citep{darkreading2023}.
Overall, downstream practitioners show little strong intention to fix vulnerabilities during the attack window from the time of issue report publication to patch release, and they may not take effective actions even when they are aware of the risk of being attacked.

During the attack window, \toolname provides practitioners with a range of options to understand potential risks in advance.
When an issue report is initially published, tools such as \toolname or MemVul can be used to make a preliminary assessment based on the description in the issue report and alert repository maintainers or downstream software developers about the potential presence of a vulnerability (if the assessment is positive).
This provides them with time to take defensive measures. Once the fix commit becomes available, \toolname can perform further analysis to assist downstream software developers in identifying security patches (because they need to identify vulnerability-fixing commits among a large number of commits.
).
Finally, after repository maintainers associate the fix commit with the issue report (if this step is taken), \toolname can further confirm the relevance of the issue report and its resolution to security vulnerabilities.

In addition, recommendations have been proposed to timely sense patches committed to the codebase in a timely sense~\citep{pan2024unveil}.

\subsection{Reasons for misclassification of vulnerability type classification task}
To investigate the reasons for misclassification, we analyzed the samples that were incorrectly predicted by \toolname. One type of misclassification occurs at the first layer of the CWE category, where \toolname incorrectly identifies the parent CWE categories. This is the most frequent type of error, accounting for 90\% of all misclassifications. Such misclassification can be attributed to the limited number of samples in certain CWE categories, which prevents the model from learning more discriminative features. Additionally, the presence of descriptions in bug reports that resemble other CWE categories may also lead to misjudgments by \toolname. Another type of misclassification occurs at the second layer, where \toolname correctly identifies the sample as belonging to the CWE-664 subclass but misclassifies the specific subclass. This type of misclassification is primarily caused by the unclear boundaries between each CWE category (due to the complexity and diversity of software vulnerabilities, there may be overlaps or ambiguous boundaries between certain categories), which is further influenced by the descriptions in bug reports.

\section{Threats to Validity}

\subsection{Internal validity}
The primary threats stem from the handling of a limited length of data.
The extraction of buggy code in the dataset is straightforward, involving the entire fragments of functions and classes where the patch code is located, including deletions and additions.
However, the handling capacity of BERT or CodeBERT for text or code length is limited, and excessively long text or code fragments may lead to the loss of important text or code segments.
Fortunately, most of the patches do not exceed the length limit.
Another concern arises from the dataset itself. 
A small number of CVEs do not have associated CWE categories, and we addressed this by manually assigning CWE numbers to them with the help of ChatGPT.
Two of the authors double-checked the labeling results in case of any bias.
There are some commits in the openEuler dataset that have not been marked as CVE patches, although we manually analyzed the results of the model predictions.
However, there may be labeling errors in the large training set, and the cost of manual analysis is high, so we do not correct the labels. Therefore, this part of the dataset is also subject to internal threats.

Another potential threat arises from the merging of CWE categories.
Due to the limited number of samples in some CWE classes, we merged several low-frequency CWE categories into higher-level composite categories following the hierarchical structure defined by MITRE.
Although this strategy improves the statistical reliability of training and evaluation, it also reduces the granularity of the vulnerability type classification task.
As a consequence, the reported classification performance may be higher than what would be observed under a fine-grained CWE taxonomy where categories are more numerous and often sparsely annotated.
Furthermore, different studies may adopt different CWE taxonomies (e.g., fine-grained CWE IDs, parent-level CWE categories, or custom merged classes), which may affect the direct comparability of results across approaches.
However, our merging strategy groups semantically related CWE categories that share similar root causes and remediation patterns.
Therefore, while the task becomes more coarse-grained, the resulting classification still reflects meaningful vulnerability families rather than arbitrary groupings.

\subsection{External validity}
The threat to external validity pertains to the generalizability of \toolname. Our dataset is constructed based on bug reports from open-source GitHub repositories. 
However, the collected security vulnerabilities and their patches may be relatively scarce compared to the real world, limiting their representativeness.
Moreover, we specifically focus on commits involving .c, .cc, .java, and .py files, which may not be representative enough.
Additionally, the model relies on various pieces of information, with the prerequisite that commits must be associated with bug reports to obtain information from bug descriptions. 
This places constraints on the model inputs.
However, ablation studies and RQ5 suggest that sacrificing some performance for fewer model inputs is feasible, especially for vulnerability identification.

%% file: conclusion.tex
 \section{Conclusion and Future Work}

In this paper, we introduced \toolname, a deep learning-based model that fuses inter-connected multi-source information from issue reports and fix commits including commit messages and patches.
Relying on the multi-source information fusion layer, \toolname is capable of mining high-level semantic information from inter-connected multi-source data and establishing potential relationships among them.
The experiments demonstrate that \toolname outperforms all state-of-the-art (SOTA) methods, validating the effectiveness of the embedding-level fusion approach.
Due to the limited number of training samples and the influence of the description of issue reports, \toolname struggles to learn the boundaries between certain CWE categories (or the distinguishing features that separate them from other categories).
In the future, we plan to further study how to better encode and utilize the information of the CWE Information to assist type classification tasks.

%% file: acknowledgement.tex
\section*{Acknowledgement}

This work was supported by the National Key R\&D Program of China No 2024YFB4506200.